\documentclass[fleqn,usenatbib,useAMS]{mnras}

\usepackage{graphicx}	
\usepackage{amsmath}	
\usepackage{amssymb}	
\usepackage{multicol}        
\usepackage{bm}		
\usepackage{pdflscape}	
\usepackage{xcolor}
\usepackage{adjustbox}

\newcommand{\be}{\begin{equation}}
\newcommand{\ee}{\end{equation}}
\newcommand{\ba}{\begin{eqnarray}}
\newcommand{\ea}{\end{eqnarray}}

\newcommand{\fracb}[2]{\left(\frac{#1}{#2}\right)}

\newcommand\restr[2]{{
  \left.\kern-\nulldelimiterspace 
  #1 
  \vphantom{\big|} 
  \right|_{#2} 
  }}
  
\definecolor{blazeorange}{rgb}{1.0, 0.4, 0.0}
\definecolor{seagreen}{rgb}{0.18, 0.55, 0.34}
\definecolor{rufous}{rgb}{0.66, 0.11, 0.03}
\definecolor{royalfuchsia}{rgb}{0.79, 0.17, 0.57}
\definecolor{scarlet}{rgb}{1.0, 0.13, 0.0}
\definecolor{royalpurple}{rgb}{0.47, 0.32, 0.66}
\definecolor{darkblue}{rgb}{0, 0, 0.66}

\usepackage[T1]{fontenc}
\usepackage{ae,aecompl}
\usepackage{newtxtext,newtxmath}

\usepackage{ulem}

\title[Robust Features of Off-Axis GRBs]{Robust Features of Off-Axis Gamma-Ray Burst Afterglow Lightcurves}

\author[Beniamini, Gill, \& Granot]{
Paz Beniamini$^{*1,2}$\thanks{E-mail: paz.beniamini@gmail.com},
Ramandeep Gill$^{*3,2}$\thanks{E-mail: r.gill@irya.unam.mx},
Jonathan Granot$^{1,2,4}$\thanks{E-mail: granot@openu.ac.il}
\\
$^{1}$Department of Natural Sciences, The Open University of Israel, P.O Box 808, Ra'anana 4353701, Israel\\
$^{2}$Astrophysics Research Center of the Open university (ARCO), The Open University of Israel, P.O Box 808, Ra'anana 4353701, Israel\\
$^{3}$Instituto de Radioastronomía y Astrofísica, Universidad Nacional Autónoma de México, Antigua Carretera a Pátzcuaro \# 8701, \\
Ex-Hda. San José de la Huerta, Morelia, Michoacán, México C.P. 58089 \\
$^{4}$Department of Physics, The George Washington University, Washington, DC 20052, USA
}

\pubyear{2022}
\begin{document}
\label{firstpage}
\pagerange{\pageref{firstpage}--\pageref{lastpage}}
\maketitle

\begin{abstract}
The ultra-relativistic outflows powering gamma-ray bursts (GRBs) acquire angular structure through their interaction with external material. They are often characterized by a compact, nearly uniform narrow core (with half-opening angle $\theta_{c,\{\epsilon,\Gamma\}}$) surrounded by material with energy per unit solid angle ($\epsilon=\epsilon_c\Theta_{\epsilon}^{-a}$, where $\Theta_{\{\epsilon,\Gamma\}}=[1+\theta^2/\theta_{c,\{\epsilon,\Gamma\}}^2]^{1/2}$) and initial specific kinetic energy ($\Gamma_0-1=[\Gamma_c-1]\Theta_\Gamma^{-b}$) declining as power laws. Multi-wavelength afterglow lightcurves of off-axis jets (with viewing angle $\theta_{\rm obs} > \theta_c$) offer robust ways to constrain $a$, $b$ and the external density radial profile ($\rho\propto R^{-k}$), even while other burst parameters may remain highly degenerate. We extend our previous work on such afterglows to include more realistic angular structure profiles derived from three-dimensional hydrodynamic simulations of both long and short GRBs (addressing also jets with shallow angular energy profiles, whose emission exhibits unique evolution). We present afterglow lightcurves based on our parameterized power-law jet angular profiles for different viewing angles $\theta_{\rm obs}$ and $k=\{0,1,2\}$. We identify a unique evolutionary power-law phase of the characteristic synchrotron frequencies ($\nu_m$ and $\nu_c$) that manifests when the lightcurve is dominated by emission sensitive to the angular structure of the outflow. We calculate the criterion for obtaining single or double peaked light-curves in the general case when $\theta_{c,\Gamma}\neq\theta_{c,\epsilon}$. We emphasize how the shape of the lightcurve and the temporal evolution of $\nu_m$ and $\nu_c$ can be used to constrain the outflow structure and potentially distinguish between magnetic and hydrodynamic jets. 
\end{abstract}



\def\thefootnote{*}\footnotetext{These authors contributed equally to this work}
\section{Introduction}
\label{sec:Intro}
The understanding that gamma-ray bursts (GRBs) originate from ultra-relativistic jets emerged in stages. First, it became clear that the detection of sub-MeV prompt emission photons from cosmological GRBs requires the emission region to be moving towards us at bulk Lorentz factors (LFs) $\Gamma\gg1$. Second, because of relativistic beaming we observe only a region of angle $\lesssim\,$1$/\Gamma$ around our line of sight, so the properties of this emission would be unchanged as long as the outflow's half-opening angle exceeds $1/\Gamma$ \citep{Paczynski-86,Goodman-86,Woods1995}.  Moreover, during this early phase the jet cannot significantly spread sideways since its angular causal size is also $\sim\,$1$/\Gamma$. This opened up the possibility of narrowly beamed jets (thus reducing the large energy requirements), the evidence for which came in the form of jet breaks, as predicted for an outflow with sharp edges \citep{Rhoads-97,Rhoads1999,SPH1999}. Such a jet break manifests as an achromatic steepening of the afterglow lightcurve corresponding to the observer time at which $\Gamma\approx \theta_{\rm c}^{-1}$ (where $\theta_{\rm c}$ is the opening angle of the jet core). It was first  detected in GRB 990510 \citep{Harrison1999,Stanek1999} and has subsequently been detected in many other GRBs, thus confirming their jetted structure.

The discovery of jet breaks led to several authors proposing a different interpretation of them based on outflows with angular structure \citep{Lipunov+01,Rossi+02,Zhang-Meszaros-02}. In particular, \citet{Rossi+02} have proposed a specific, ``universal'', structure in which both the kinetic and $\gamma$-ray emitted energy per unit solid angle decline as $\epsilon\propto \theta^{-2}$ for polar angles larger than the core angle ($\theta>\theta_c$). In such a scenario, jet breaks still exist, but they now correspond to $\Gamma \sim \theta_{\rm obs}^{-1}$ (where $\theta_{\rm obs}$ is the observer viewing angle w.r.t the jet symmetry axis). This specific model for the angular distribution of energy is discouraged by several observational considerations, including burst energetics and rates \citep{NGG2004}, the shape of the GRB afterglow lightcurve before the jet break \citep{Kumar-Granot-03} (depending also on the Lorentz factor angular distribution) and the polarization around the jet break time \citep{Rossi+04}. Nevertheless, it drew the community's attention to the importance of studying the angular structures of GRB jets beyond their cores, and finding means by which these structures and $\theta_{\rm obs}/\theta_c$ could be constrained 
by observations.

\cite{Beniamini-Nakar-19} have found that in most cosmological GRBs, the $\gamma$-ray emissivity drops rapidly beyond their cores (either because the jet structure declines steeply, or because the efficiency of $\gamma$-ray production drops rapidly at such angles). As a result, typical GRBs are likely viewed at angles close to, or within, their jet's core (a similar result was also obtained by \citet{Gill+20} using compactness arguments in angular structured outflows). Still, jets viewed from angles slightly beyond their core, may hold the key towards explaining some of the most intriguing phenomena routinely observed in X-ray afterglow lightcurves: plateaus \citep{Eichler-Granot-06,BDDM2020} and X-ray flares \citep{Duque21}. For GRB jets that are significantly misaligned from the observer (which far outnumber the population of aligned jets since the jet orientation relative to the observer is random), the prompt emission is much fainter and may easily be missed. Indeed,  in the first gravitational wave (GW) detected neutron star merger GRB \citep{Abbott+17-GW170817-GRB170817A}, GRB 170817A, the isotropic-equivalent $\gamma$-ray energy release was several orders of magnitude fainter than observed in typical short GRBs (sGRBs). In that instance, it was only thanks to the GW detection that extensive follow-up observations across the electromagnetic spectrum were done, and the afterglow was detected. These observations, and specifically the shallow rise to the lightcurve peak at $t_{\rm pk}\sim150\,$days and its subsequent decline \citep[e.g.,][]{Margutti2018,DAvanzo2018,Troja2018,LMR2018,Hajela2019,Lamb2019,Makhathini+21} as well as the movement of the flux centroid over time \citep{Mooley2018,Ghirlanda2019}, enabled the determination that the underlying GRB had an energetic core, misaligned from Earth, and a steep angular structure beyond that core \citep[e.g.,][]{GG2018,Lazzati+18,Margutti2018,Gill+19,Ghirlanda2019,Hotokezaka+19,Troja2019,Wu-MacFadyen-19,Ryan+20,BGG2020,Nakar-Piran-21,Nathanail+20,Nathanail+21}. In the near future, more GW detected sGRBs should be observable \citep{BPBG2019,Duque2019,Gottlieb2019} and it will be possible to examine their outflow structures in much more detail.

Beside GW detected GRBs, misaligned GRBs may also be detectable through ``orphan afterglows" \citep{Rhoads-97,Nakar2002} in bursts where the prompt emission was missed but the afterglow was detectable as a radio, optical, or X-ray transient. The emergence of increasingly sensitive all-sky surveys in recent years, such as ZTF and Pan-STARRS has already led to some orphan afterglow candidates \citep[e.g.,][]{Ho2020} and is likely to result in many more events in the years to come (e.g. thanks to the upcoming Vera Rubin Observatory). Due to their greater energies and comparable local intrinsic rates\footnote{Above a luminosity $L>10^{50}\mbox{erg s}^{-1}$, the observed local rate is $1.3f_{\rm b}^{-1}\mbox{Gpc}^{-3}\mbox{ yr}^{-1}$ ($ 2.1f_{\rm b}^{-1}\mbox{Gpc}^{-3}\mbox{ yr}^{-1}$) for long (short) GRBs \citep{WP2010,WP2015} where $f_{\rm b}$ is the jet beaming factor (the overall rate of sGRBs is estimated to be at least twice as large as the number quoted above, since it is estimated that $L_{\rm min}\lesssim 5\times 10^{49}\mbox{erg s}^{-1}$ for sGRBs). While still somewhat uncertain, $f_{\rm b}$ is constrained by jet break time observations (mainly in long GRBs) and, in short GRBs, by comparison to the LIGO-VIRGO measured NS merger rate \citep{BPBG2019}. These comparisons suggest $f_{\rm b}$ is not drastically different between long and short GRBs. Due to the harder luminosity function of lGRBs (as compared to sGRBs), the rate of lGRBs relative to sGRBs increases with limiting luminosity. Thus, depending on the limiting sensitivity of a given survey, the number of off-axis jet discoveries might be dominated by lGRBs, rather than sGRBs.}, long GRBs (lGRBs), rather than sGRBs, are likely to dominate this channel of misaligned GRB jet discoveries.

The expectation of a growing sample of afterglows from misaligned GRB jets lead us in a previous work (\citealt{BGG2020}, henceforth BGG20) to explore what can be learned about the outflow structure and $\theta_{\rm obs}$ from the shape of the lightcurve. In the present work, we expand on BGG20 in a few ways. First, we extend the analysis to GRB jets with a shallow energy profile (which lead to qualitatively different afterglows) and/or that are propagating in a general external density profile. The latter is done mainly to accommodate lGRBs that are expected to reside in the wind-like environments produced by their progenitor stars. Secondly, we consider structures that are motivated by hydrodynamical GRB jet simulations (allowing for example for the core of the Lorentz factor profile to be wider than that of the energy per unit solid angle). We focus on robust features of numerical jet structures and discuss how they would 
manifest in the afterglow observations. Finally, we demonstrate the unique temporal evolution of the synchrotron characteristic frequencies that can be inferred for off-axis jets and how it, in addition to the afterglow lightcurve, can provide new constraints on the outflow structure and viewing geometry.

The paper is organized as follows. In \S \ref{sec:model}, we briefly outline the misaligned structured jet afterglow model parameters and stress the new additions relative to our model as described in BGG20. We then show in \S \ref{sec:frequencies} how misaligned structured jets, lead to unique temporal evolutions of the characteristic synchrotron frequencies and describe how this can provide novel constraints on the energy structure, the external density profile and the power-law index $p$ of the accelerated electrons' energy distribution. In \S \ref{sec:shallowjet} we apply our modelling to jets with shallow energy structures (motivated by certain hydrodynamic GRB jet simulations) and show they can be clearly distinguished from steeper structures. In \S \ref{sec:Jetstructure} we describe features of long and short GRB jet structures as informed by hydrodynamical simulations. We then explore the implications for the observed lightcurves of misaligned sGRBs in \S \ref{sec:sGRBs} and lGRBs in \S \ref{sec:lgrbs}. We conclude in \S \ref{sec:discuss}.

\section{The model}
\label{sec:model}
The modelling used in this paper largely follows the one described in our earlier work (BGG20) with some minor changes to accommodate slightly more general jet structures that will be described below. We briefly outline some of the key features below, and refer the reader to BGG20 for more details.

The energy per unit solid angle and initial (i.e., before deceleration) Lorentz factor (LF) profiles as a function of the angle $\theta$ from the jet symmetry axis are described by smoothly broken power laws:
\begin{equation}\label{eq:PLJ}
        \frac{\epsilon(\theta)}{\epsilon_{\rm c}} = \Theta_{\epsilon}^{-a}~,\quad\frac{\Gamma_0(\theta)-1}{\Gamma_{\rm c,0}-1} = \Theta_{\Gamma}^{-b}~,
        \quad\quad\Theta_X \equiv \sqrt{1+\displaystyle\fracb{\theta}{\theta_{\rm c,X}}^2}\ ,
\end{equation}
where, motivated by results of numerical simulations, we have allowed for the core opening angles for the $\epsilon$ and $\Gamma$ profiles ($\theta_{\rm c, \epsilon},\theta_{\rm c,\Gamma}$) to differ from each other.  Typically, we expect $1\lesssim \theta_{\rm c,\Gamma}/\theta_{\rm c, \epsilon}\lesssim 2$, i.e. core of the LF angular profile is slightly wider than that of the energy.

As in BGG20, we also define the relative viewing angle $q\equiv \theta_{\rm obs}/\theta_{\rm c, \epsilon}$ and the compactness of the core $\xi_{\rm c}\equiv(\Gamma_{\rm c,0}\theta_{\rm c,\epsilon})^2$. Since $\Gamma$ can have a wider core than $\epsilon$, it is useful to also define $\bar{\xi}_{\rm c}\equiv(\Gamma_{\rm c,0}\theta_{\rm c,\Gamma})^2=\xi_{\rm c} (\theta_{\rm c, \Gamma}/\theta_{\rm c,\epsilon})^{2}$. In BGG20 we have shown that the type of observed lightcurve (e.g. single or double peaked) depends on the value of $\theta_{\rm obs}/\theta_*$. If $\theta_{\rm obs}\gg\theta_*$, then angles much lower than $\theta_{\rm obs}$ are observable from the earliest stages of the emission. This means that after an initial phase (i) in which the material dominating the early flux (located at $\theta_{\rm F,0}\ll\theta_{\rm obs}$) coasts and finally decelerates, the flux becomes gradually dominated by ever decreasing latitudes (ii) that are coming into the observer's view 
(i.e. the beaming cone of the emitting material at these latitudes now contains the observer's line-of-sight). This lasts until the jet's core becomes visible to the observer, after which point the flux declines (iii) in a manner similar to an on-axis jet viewed after the jet break time.  The result is therefore a single peaked lightcurve. If instead $\theta_{\rm obs}\ll\theta_*$, then the material dominating the initial flux is located at $\theta_{\rm F,0}\approx\theta_{\rm obs}$ and decelerates while it is still highly beamed towards the observer (i.e. well before the emission from angles $\theta\ll\theta_{\rm obs}$ becomes visible). As a result, there is an extra phase in the lightcurve, between steps (i) and (ii) mentioned above, in which the flux declines as it would for an on-axis jet post-deceleration but before the jet break (as it is dominated by $\theta\simeq\theta_F\approx\theta_{\rm obs}$ since $\theta_{\rm obs}\Gamma(\theta_{\rm obs})\gg1$). This lasts until such times for which emission from material below $\theta_{\rm F,0}\approx\theta_{\rm obs}$ comes into view of the observer, ($\theta_{\rm obs}\Gamma(\theta_{\rm obs})\sim1$) and phase (ii) begins.
The overall result is a double peaked lightcurve, with the first peak corresponding to deceleration of material at $\theta_{\rm F,0}\approx\theta_{\rm obs}$ and the second to the jet's core coming into view of the observer. 

The modified value of $\theta_*$, accounting for the possibility of $\theta_{\rm c,\Gamma}/\theta_{\rm c,\epsilon}\geq 1$, is given by
\begin{equation}
\label{eq:thetastar}
    \theta_*=\theta_{\rm c,\Gamma}\,\bar{\xi}_c^{\,\frac{1}{2(b-1)}}\;.
\end{equation}
The wider LF core leads to a modification in the time of the first peak (that exists for $\theta_{\rm obs}\ll \theta_*$), such that
\begin{equation}
    \tilde{t}_{\rm 1pk}=\bigg(\frac{\theta_{\rm c,\epsilon}}{\theta_{\rm c,\Gamma}}\bigg)^{\frac{2b(4-k)}{3-k}} (1+q_{\rm F,0}^2)^{\frac{2b(4-k)-a}{2(3-k)}}
\end{equation}
where $q_{\rm F,0}\equiv \theta_{\rm F,0}/\theta_{\rm c,\epsilon}$, $\theta_{\rm F,0}$ is the angle from which emission initially dominates 
the observed flux, $\tilde{t}$ denotes the apparent time in units of the core's deceleration time, and $k$ relates the external density to the radius, 
i.e $\rho=AR^{-k}$. The flux at the time of the first peak is given by
\begin{equation}
    F_{\rm 1pk}=F_{\rm pk}q^{\frac{8(k-4)-a[k(p+5)-4(p+3)]}{4(k-4)}} \bigg(\frac{t_{\rm 1pk}}{t_{\rm pk}}\bigg)^{\frac{3kp-5k-12p+12}{{4(4-k)}}}
\end{equation}
where $t_{\rm pk},F_{\rm pk}$ are the time and flux of the main (i.e. second) peak in the lightcurve. This expression is in agreement with Eq.\,(41) of BGG20 in the appropriate limit of $k=0$.

Finally, we provide the general form of the double peaked (cases 1A, 2, and 3 of BGG20), $F_{\rm dbl}$, and the single-peaked (case 1B), $F_{\rm sngl}$, lightcurves:
\begin{eqnarray}
\label{eq:dbl}
& F_{\rm dbl}=F_{\rm 1pk}\bigg(\frac{t}{t_{\rm 1pk}}\bigg)^{\alpha_r}\bigg[\frac{1+(t/t_{\rm 1pk})^2}{2}\bigg]^\frac{\alpha_d-\alpha_r}{2}\quad\quad\quad
\nonumber  \\ & \quad\quad + F_{\rm pk}\bigg(\frac{t}{t_{\rm pk}}\bigg)^{\alpha}\bigg[\frac{1+(t/t_{\rm pk})^2}{2}\bigg]^{\frac{\alpha_f-\alpha}{2}}e^{-(t_{\rm 1pk}/t)^2}
\end{eqnarray}
\begin{eqnarray}
\label{eq:sngl}
 F_{\rm sngl}\!=\!F_{\rm pk} \bigg[1\!+\!\bigg(\frac{t}{t_{\rm dec}(\theta_{F,0})}\bigg)^{-4}\bigg]^{\frac{\alpha\!-\!\alpha_r}{4}} \bigg(\frac{t}{t_{\rm pk}}\bigg)^{\alpha}\bigg[\frac{1\!+\!(t/t_{\rm pk})^2}{2}\bigg]^{\frac{\alpha_f\!-\!\alpha}{2}}
\end{eqnarray}
where $t_{\rm dec}(\theta)$ is the deceleration time of material at a given $\theta$. The exponential cutoff in the second line of 
Eq.\,(\ref{eq:dbl}) ensures that the flux from low latitudes of the jet, initially beamed away from view, only contributes after the 
corresponding parts of the jet become visible to the observer (this factor was not explicitly written in BGG20, as in the parameter 
regime considered in that work, the extension of the $t^{\alpha}$ never dominated the emission at $t<t_{\rm 1pk}$).

\section{Evolution of characteristic frequencies for structured jets}
\label{sec:frequencies}
A unique phase in the temporal evolution of afterglows from structured jets (that does not appear in afterglows of jets viewed along their cores) occurs while the emission progressively becomes dominated by ever decreasing latitudes as the polar angle of material dominating the observed flux, $\theta_{\rm F}$, decreases with time. We dub this phase here as the `angular structure dominated emission' (ASDE) phase. For completeness, we also dub the early stage where the emission is dominated by material along the line of sight, as `LoS dominated emission' (LoSDE). Finally, we dub the latter stage, in which the (off-axis) core dominates the emission as `core-dominated emission' (CDE). The ASDE phase leads to a unique temporal evolution of the flux (see \citealt{GG2018}; BGG20). As we show here it also leads to a unique temporal evolution of the characteristic frequencies in the synchrotron spectrum. Interestingly, the rate of change of these frequencies within the different temporal power-law segments is independent of the structure of the jets, i.e. it has no dependence on the angular structure parameters $a$ and $b$ (this does not hold for the `shallow angular structure dominated emission' (sASDE) phase that appears in jets with a sufficiently shallow energy profile, which is discussed in \S\,\ref{sec:shallowjet}). While this is not useful for directly probing the jet structure from the characteristic frequency evolution, it also means that the evolution is robust. Therefore, observing the unique way in which these frequencies evolve with time during the ASDE and prior/later phases, could provide a strong test for the existence of the jetted structure in a given observed event. 

To derive the temporal profile of the critical synchrotron frequency of minimal energy electrons ($\nu_m$) and of electrons that radiatively cool on the dynamical timescale ($\nu_c$), we consider their dependence on radius $R$ and bulk LF $\Gamma$. In the bulk comoving frame, the LF of electrons emitting at the cooling break frequency $\nu_c$ is $\gamma_{\rm c}\propto \Gamma B'^{-2}R^{-1}$, where $B'\propto \Gamma R^{-k/2}$ is the comoving magnetic field (all primed quantities are expressed in the comoving frame hereafter). The cooling frequency in this frame is $\nu_{\rm c}'\propto B' \gamma_{\rm c}^2\propto \Gamma^{-1}R^{\frac{3k}{2}-2}$. Transforming to the observer frame and using $R\,\propto\,\Gamma^2 t$ with $t$ being the apparent time, as appropriate for an ultra-relativistic jet in which the line-of-sight of an (on-beam) observer lies within the $1/\Gamma$ beaming cone centered on the direction of motion of the emitting material\footnote{Recall that this is true by definition for material at $\theta_{\rm min}(t)$ and that as shown by BGG20, the angle dominating the observed flux, $\theta_{\rm F}$ is asymptotically proportional to $\theta_{\rm min}$ (see Fig.~1 of that work).}, we obtain $\nu_{\rm c}\propto(\Gamma^2 t)^{(3k-4)/2}$. Recalling the definition of $\theta_{\rm min}$, $\Gamma(\theta_{\rm min},t)\equiv (\theta_{\rm obs}-\theta_{\rm min})^{-1}$ and taking $\theta_{\rm c,\epsilon},\theta_{\rm c,\Gamma}\ll \theta_{\rm min}\ll \theta_{\rm obs}$ in the ASDE regime we obtain $\Gamma\approx \theta_{\rm obs}^{-1}=\,$const during this evolution. As a result we find $\nu_{\rm c}\propto t^{(3k-4)/2}$. Similarly, $\gamma_{\rm m}\propto \Gamma$ leading to $\nu_{\rm m}\propto \Gamma^4 R^{-k/2}$. Taking $\Gamma$ once more to be constant during the ASDE, we obtain $\nu_{\rm m}\propto t^{-k/2}$. The evolution of both $\nu_{\rm m},\nu_{\rm c}$ are therefore the same as those found for a pre-deceleration ultra-relativistic jet viewed on-axis. In these regimes both frequencies depend solely on the external density parameter $k$. A similar derivation can be used to find the time evolution of $\nu_{\rm m},\nu_{\rm c}$ in other phases of the emission. For example, for an on-axis GRB with $\Gamma\gg 1$ observed after the deceleration break, we have $\Gamma \propto E^{1/2} R^\frac{k-3}{2}$, leading to $\nu_{\rm m}\propto E^{1/2}t^{-3/2}, \nu_{\rm c}\propto (Et)^\frac{3k-4}{8-2k}$ (this holds for the LoSDE phase where $E=E_{\rm k,iso}(\theta_{\rm obs})$).

The temporal evolution of $\nu_{\rm m}$ and $\nu_{\rm c}$ in the different phases of the GRB is summarized in Table\,\ref{tab:numnuc}. Observation of this type of evolution of the critical frequencies in real data could be used as a way to test that a given jet is indeed being viewed off-axis.\footnote{While the characteristic frequencies evolve the same as in the pre-deceleration phase, the flux does not, so the two phases are not degenerate.} Furthermore, it can be used as a way to measure the value of $k$ and determine the nature of the external medium encountered by the outflow. 

The above derivation relies on the fact that more internal parts of the jet will eventually dominate the observed light-curve, once they come into view of the observer (i.e. the observer lies within the $1/\Gamma$ beaming cone of material emitting at $\theta<\theta_{\rm obs}$). In such a situation, the emission progresses over time from LoSDE to ASDE and finally CDE. However, jets with very shallow angular profiles do not follow this evolution, and for those the emission will be LoSDE until late times (when they eventually become dominated by progressively {\it increasing} angles, dubbed here as the sASDE phase). Such jets have many unique temporal properties, and they are discussed in more detail in \S\,\ref{sec:shallowjet}.

\begin{table}
    \centering
    \begin{tabular}{c|c|c}
    \hline
         phase & $d\log\nu_{\rm m}/d\log t$ & $d\log \nu_{\rm c}/d\log t$ \\
        \hline
        $\Gamma\gg1$, pre-deceleration, LoSDE & $-k/2$ & $3k-4\over 2$ \vspace{0.2cm}\\
        $\Gamma\gg1$, post-deceleration, LoSDE & $-3/2$ & $3k-4\over 8-2k$ \vspace{0.2cm}\\
        ASDE & $-k/2$ & $3k-4\over 2$ \vspace{0.2cm}\\
        $\Gamma\gg1$, post-jet break - CDE (NLS)  & $-3/2$ & $3k-4\over 8-2k$ \vspace{0.2cm}\\
        $\Gamma\gg1$, post-jet break - CDE (MLS)  & $-2$ & $0$ \vspace{0.2cm}\\
        $\Gamma\beta < 1$ - CDE & $4k-15 \over 5-k$ & $2k-1\over 5-k$ \vspace{0.2cm}\\
        $\Gamma\beta < 1$, Deep Newtonian - CDE & $-{3 \over 5-k}$ & $2k-1\over 5-k$ \vspace{0.2cm}\\
        sASDE & ${-{a(k-3) \over 2(a+2k-8)}}-\frac{3}{2}$ & $(a-2)(3k-4)\over 2(a+2k-8)$ \vspace{0.2cm}\\
        \hline
    \end{tabular}
    \caption{Temporal evolution of characteristic frequencies in different phases of an off-axis GRB afterglow. For the post-jet break evolution we provide solutions for two limiting cases, i.e. no lateral spreading (NLS) and maximal lateral spreading (MLS). The deep Newtonian regime is realized when in addition to the bulk velocity becoming non-relativistic, only a small (and decreasing with decreasing $\beta$) fraction of the electrons can maintain relativistic thermal Lorentz factors \citep{Granot+06}. We note that for $\Gamma\beta<1$, the approximation of no lateral spreading must break down. However, the blast-wave eventually approaches the spherical 
    Sedov-Taylor solution, and as a result the scalings of $\nu_m,\nu_c$ reported for this case still hold.}
    \label{tab:numnuc}
\end{table}

\begin{figure*}
    \centering
    \includegraphics[width=0.45\textwidth]{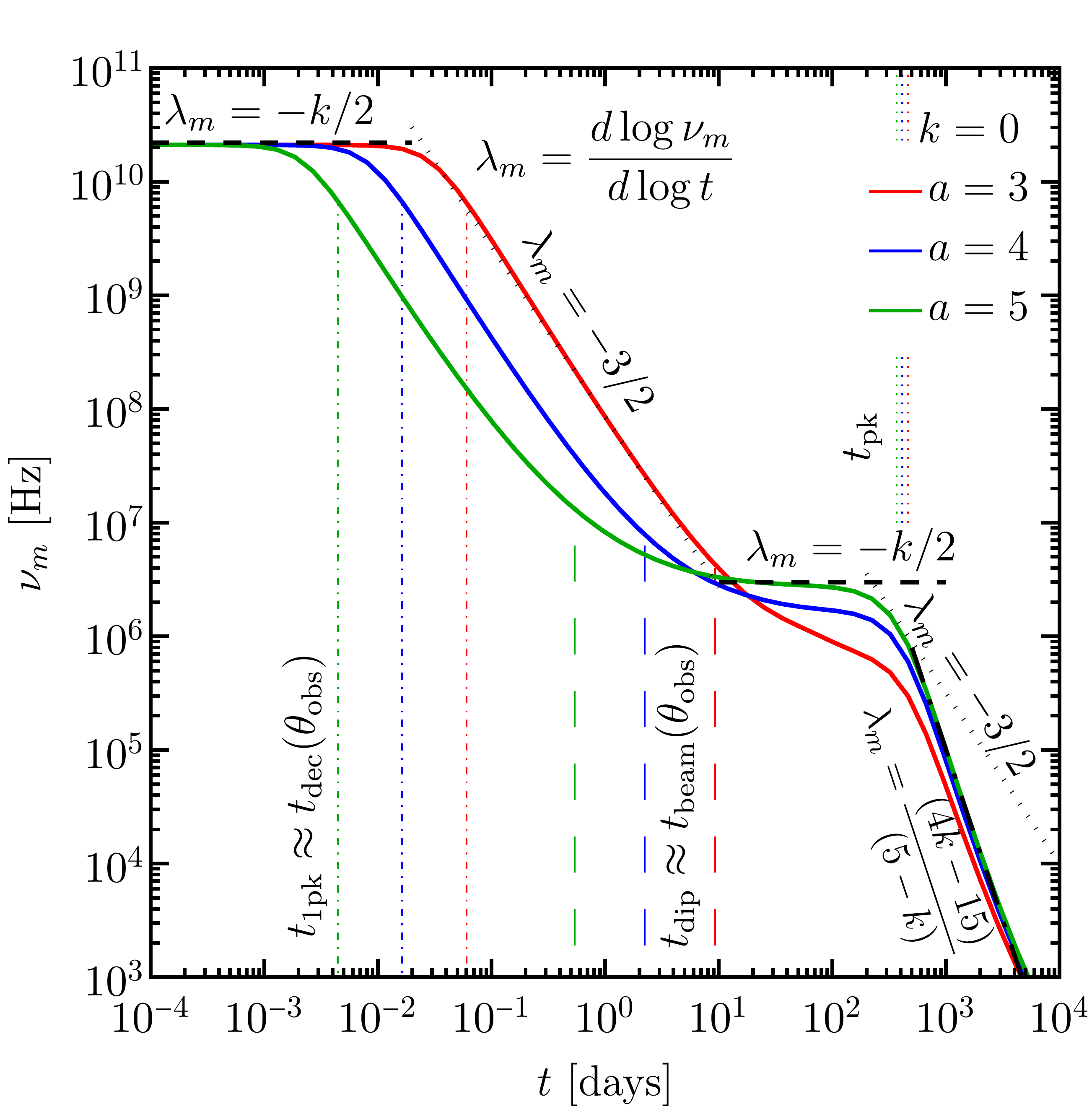}\quad\quad
    \includegraphics[width=0.45\textwidth]{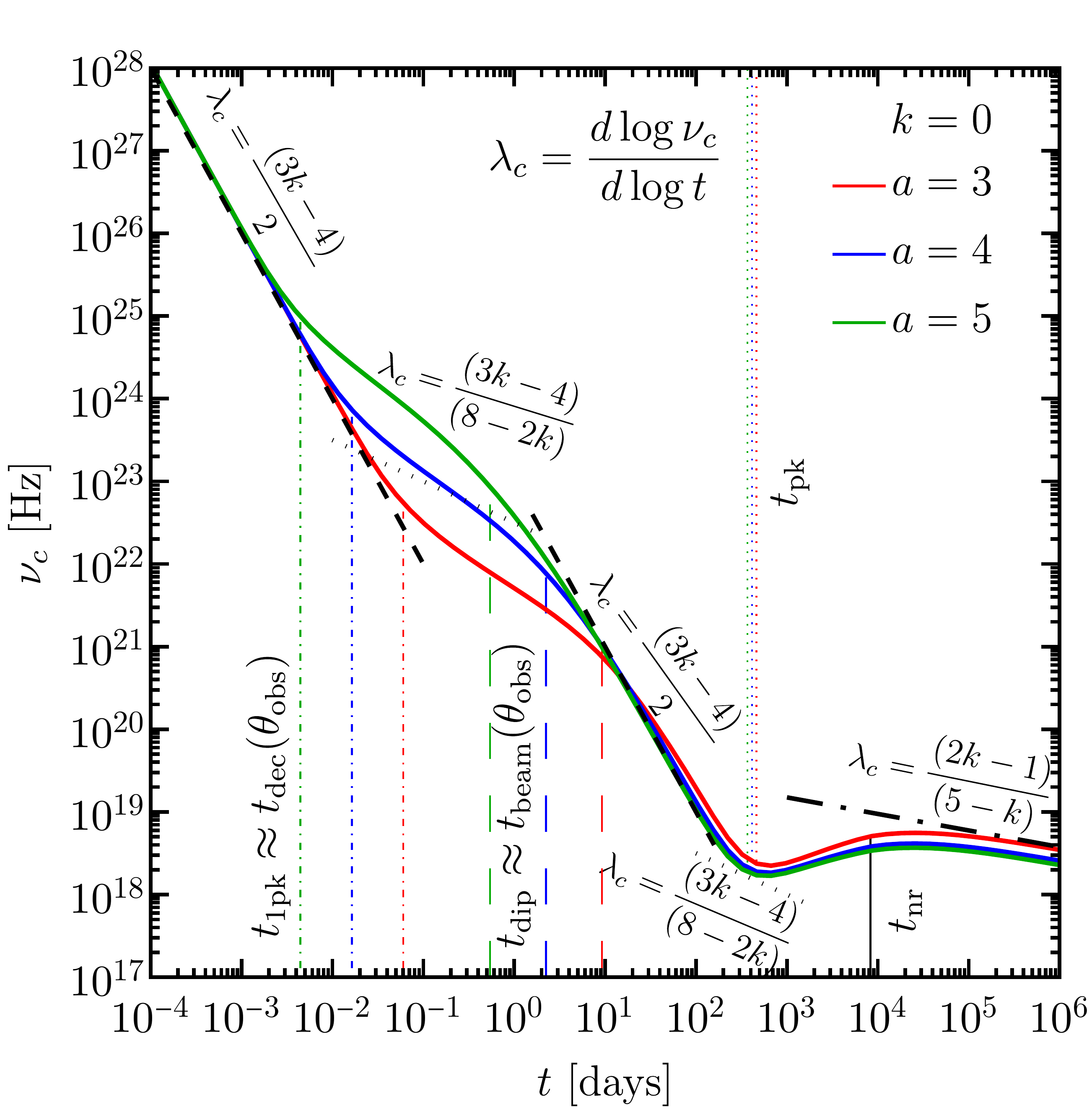} \\
    \includegraphics[width=0.45\textwidth]{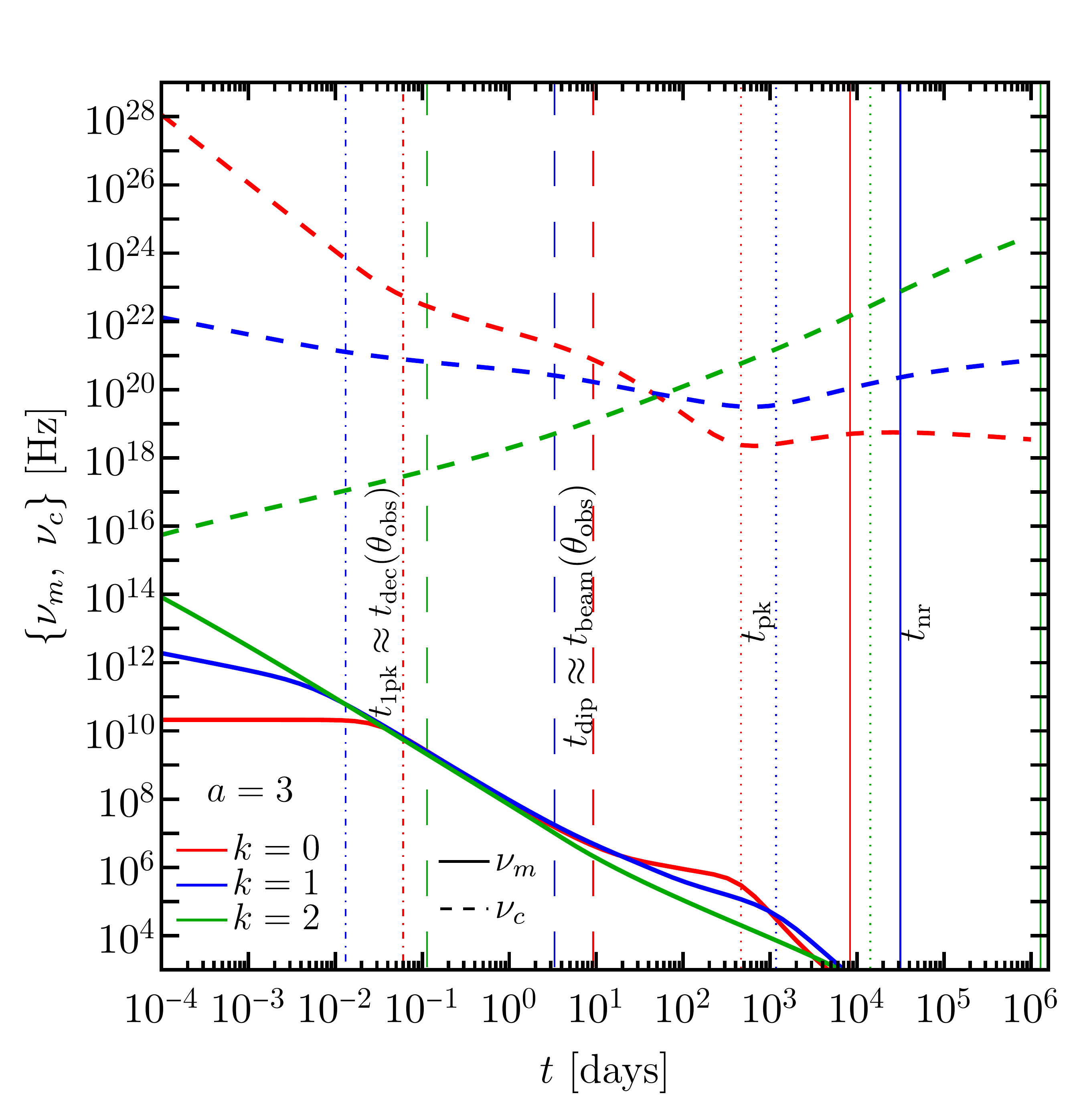}
    \caption{Temporal evolution of critical synchrotron frequencies for jets with no lateral spreading (NLS) and different power-law indexes for the jet's energy angular profile ($a$) and the external density radial profile ($k$). The dashed, dotted, and dash-dotted black lines indicate the asymptotic trend from analytical calculations (see text for more details). The model parameters used for this plot are: $b=1.25$, $\epsilon_c=5\times10^{49}\,{\rm erg\,sr}^{-1}$, $\Gamma_{c,0}=2\times10^3$, $\theta_c=0.01$, $\theta_{\rm obs}=0.5$, $n(R_0)=10^{-5}\,{\rm cm}^{-3}$, $R_0=10^{18}\,{\rm cm}$, $\epsilon_e=0.05$, $\epsilon_B=0.05$, $p=2.2$, 
    $\xi_{e,0}=1$. These parameters were chosen to have sufficient dynamical range for the different power-law segments in the figure. The large range of frequencies and timescales shown are only for demonstrating the asymptotic trend of the break frequencies obtained for a set of fiducial parameters. The different characteristic timescales are shown with different line traces for different $k$, where $t_{\rm 1pk}$ (dot-dashed) is the peak time for the first peak, $t_{\rm dip}$ (dashed) is the time for the dip between the two lightcurve peaks, $t_{\rm pk}$ (dotted) is the time for the second peak when the jet core becomes visible, and $t_{\rm nr}$ (solid) is the non-relativistic transition time for the jet core. The rise in the slope of $\nu_c$ just before $t\sim t_{\rm nr}$ is not captured by the asymptotic 
    slopes that are only valid away from the relativistic to non-relativistic transition.}
    \label{fig:nu_m-nu_c-Vs-t}
\end{figure*}

In Fig.~\ref{fig:nu_m-nu_c-Vs-t} we present the temporal evolution of synchrotron critical frequencies ($\nu_m$ and $\nu_c$) for different power-law indexes for the jet's energy angular profile ($a$) and the external density radial profile ($k=0,1,2$). In the top-left and top-right panels, it can be seen that at very early times, for $t\ll t_{\rm dec}(\theta_{\rm obs})$, both $\nu_m$ and $\nu_c$ are independent of the energy angular structure (but they do depend on the LF profile). This results from the fact that the emission is still dominated by that arising from $\theta\approx\theta_{\rm obs}$, in which case it cannot be affected by the jet's angular structure but only by $\Gamma(\theta_{\rm obs})$ and $t$. Furthermore, at this point the material at $\theta_{\rm obs}$ has not decelerated yet, which means that $\Gamma(\theta_{\rm obs})=\Gamma_0(\theta_{\rm obs})$ and the shocked material has smaller energy and is unaware of the energy of the ejecta ($E_{\rm k,iso}(\theta_{\rm obs})$) that acts like a piston. Therefore, when all other parameters are the same, the normalization of $\nu_m$ and $\nu_c$ becomes independent of the energy angular structure. For $t_{\rm dec}(\theta_{\rm obs})< t < t_{\rm beam}(\theta_{\rm obs})$ the normalization of the critical frequencies are different for different $a$ even though the temporal slope, which can only depend on the external density profile, is the same. The difference in normalization here is produced by the difference in isotropic-equivalent energy $E(\theta_{\rm obs})$ for the different power-law index $a$ (where the emission in this phase is the same as for a \citet{BM76} spherical self-similar solution with energy $E=E_{\rm k,iso}(\theta_{\rm obs})$). The slopes of $\nu_m,\nu_c$ during the ASDE reproduce our analytic estimates above. During this phase, the bulk LF at $\theta_F\approx\theta_{\rm min}$ is asymptotically constant for $\theta_{\rm min}\ll\theta_{\rm obs}$, however, some slight differences in the relative normalization still remain. These are more pronounced for $\nu_m$ whereas in $\nu_c$ the normalization is very similar. For the post-jet-break evolution of the critical frequencies, the dependence on $E$ disappears since $\epsilon_c$ is the same for all the curves, and therefore the normalization for different $a$ should be very similar. The bottom panel of Fig.~\ref{fig:nu_m-nu_c-Vs-t} shows the dependence of the characteristic frequencies on $k$. Since $\nu_m(t)\propto\rho^0E_{\rm k,iso}^{1/2}$ is independent of the external density $\rho$, and $E_{\rm k,iso}(\theta_{\rm obs})$ is the same in all cases shown in this panel, all the $\nu_m(t)$ curves coincide. 

The dynamical evolution of the (locally) spherical outflow becomes non-relativistic when $\Gamma^2(\theta,R)\approx E_{\rm k,iso}(\theta)/M(R)c^2\sim1$, which yields the non-relativistic transition radius $R_{nr}(\theta)=[(3-k)E_{\rm k,iso}(\theta)/4\pi Ac^2]^{1/(3-k)}$, where $A=m_pn(R_0)R_0^k$. In this work we used a reference radius of $R_0=10^{18}\;$cm and the external number density is normalized at this radius.\footnote{We are aware that in some cases we chose a very low external density value. This was mainly done for demonstrative purposes and to have a large enough dynamical range that clearly shows the different regimes we discuss.} For an angular structured flow with a sufficiently sharp energy profile (see \S\,\ref{sec:shallowjet} for the shallow jet case), the observed flux is dominated by emission from the more energetic core ($\theta<\theta_c$) at $t>t_{\rm pk}$. The apparent time when the core becomes non-relativistic, in which case $E_{\rm k,iso}(\theta)=4\pi\epsilon_c$, can be approximated as $t_{\rm nr}\sim R_{\rm nr}/c$. This timescale is shown as a solid vertical line in Fig.~\ref{fig:nu_m-nu_c-Vs-t} for different values of $k$. 

Sufficiently deep into the Newtonian regime, when $\Gamma\beta\ll1$, eventually the fraction $\xi_e$ of electrons that cross the shock which are accelerated into a relativistic power-law energy distribution with LF $\gamma_m\leq\gamma_e\leq\gamma_M$ must drop below its initial value in the relativistic regime of $\xi_{e,0}\leq1$. The LF of the minimal energy electrons is sensitive to $\beta c$, the relative upstream to downstream velocity across the shock,
\begin{equation}\label{eq:gamma_m}
\gamma_m = \frac{\epsilon_e}{\xi_e}\left(\frac{p-2}{p-1}\right)\frac{m_p}{m_e}(\Gamma-1)
\approx \frac{\epsilon_e}{\xi_e}\left(\frac{p-2}{p-1}\right)\frac{m_p}{m_e}\frac{\beta^2}{2}\ ,
\end{equation}
where the latter approximation $(\Gamma-1)=(1-\beta^2)^{-1/2}-1\approx\beta^2/2$ holds in the non-relativistic regime in which $\beta\ll1$. Since the synchrotron power scales as $P'_{\rm syn}\propto u_e^2$, where $u_e=(\gamma_e^2-1)^{1/2}$ must be larger than unity for the electrons to be relativistic and emit synchrotron radiation, this yields that $\gamma_m\geq\sqrt{2}$. The transition into the deep Newtonian (denoted by a subscript `dn') regime, where this condition is violated for $\xi_e=\xi_{e,0}$, occurs at
\begin{equation}
\beta_{\rm dn}=\sqrt{2^{3/2}\frac{p-1}{p-2}\frac{\xi_{e,0}}{\epsilon_e}\frac{m_e}{m_p}}\approx0.22\,\sqrt{\frac{(p-1)}{3(p-2)}\frac{\xi_{e,0}}{\epsilon_{e,-1}}}\ .
\end{equation}
In this deep Newtonian regime some assumption on the shock microphysics must be varied, and in particular either $\epsilon_e$ or $\xi_e$ (or both) must vary. While this is rather poorly understood, here we follow \citet{Granot+06} who found that assuming a constant $\epsilon_e$ while varying $\xi_e$ such that $\gamma_m\approx\sqrt{2}$ remains constant in this regime provides a very good fit to the late time radio afterglow observations of the nebula produced by the outflow from the bright 27 December 2004 giant flare from the magnetar SGR\,1806-20. In particular, we adopt their parameterization where $\xi_e=\xi_{e,0}\min[1,(\beta/\beta_{\rm dn})^2]$. This scaling of $\xi_e$ with the shock velocity is taken into account for the evolution of critical synchrotron frequencies in Fig.~\ref{fig:nu_m-nu_c-Vs-t}, where $\xi_{e,0}=1$.

\section{Temporal evolution of emission from jets with shallow energy structures}
\label{sec:shallowjet}

\begin{table*}
    \centering
   \begin{adjustbox}{max width=\textwidth}
    \begin{tabular}{c|c|c|c|c|c|c|c|c|c}
    \hline
         PLS & $\beta$ & $a_{\rm cr}$ & $a_{\rm cr}|_{k=0,p=2.2}$ & $a_{\rm cr}|_{k=1,p=2.2}$ & $a_{\rm cr}|_{k=2,p=2.2}$ & $\alpha_{\rm s}$ & $\alpha_{\rm s}|_{k=0}$ & $\alpha_{\rm s}|_{k=1}$ & $\alpha_{\rm s}|_{k=2}$\\
        \hline
        D & $1/3$ & $\frac{24-6k}{13-4k}$ & 1.85  & 2 & 2.4 & $a(9-4k)+6(k-2)\over 3(a+2k-8)$ & $3a-4\over a-8$ & $5a-6\over 3(a-6)$ & $a\over 3(a-4)$\vspace{0.2cm}\\
        E & $1/3$ & $\frac{24-6k}{17-6k}$ & 1.41  & 1.64 & ---$^\dagger$ & $a(11-6k)+6k-4\over 3(a+2k-8)$  & $11a-4 \over 3(a-8)$ & $5a+2\over 3(a-6)$ & $8-a \over 3(a-4)$\vspace{0.2cm}\\
        F & $-1/2$ & $\frac{32-8k}{16-3k}$ & 2 & 1.85 & 1.6 & $a(8-3k)-2k+8 \over 4(a+2k-8)$  & $2(a+1)\over a-8$ & $5a+6\over 4(a-6)$ &$a+2\over 2(a-4)$\vspace{0.2cm}\\
        G  & $\frac{1-p}{2}$ & $\frac{32-8k}{16+4p-5k-kp}$ & 1.29 & 1.36 & 1.54 & $-{(a+6)kp+(a-2)(5k-12)-24p\over 4(a+2k-8)}$  & $3(a+2(p-1)) \over a-8$ & $a(7-p)+18p-14\over 4(a-6)$ & $(6-a)p+a-2\over 2(a-4)$ \vspace{0.2cm}\\
        H  & $-\frac{p}{2}$ & $\frac{32-8k}{12+4p-2k-kp}$ & 1.54  & 1.45 & 1.29 & $-akp-2ak+8a-6kp+4k+24p-16 \over 4(a+2k-8)$  & $2(a+3p-2)\over a-8$ & $-\frac{ap-6a-18p+12}{4(a-6)}$ & $a(p-2)-6p+4\over 8-2a$\vspace{0.2cm}\\
        \hline
    \end{tabular}
    \end{adjustbox}
    \caption{For each of the synchrotron PLS (as defined by \citealt{Granot-Sari-02}) and characterized by a spectral slope $\beta$, such that $F_{\nu}\propto \nu^{\beta}$ we provide values of $a_{\rm cr},\alpha_{\rm s}$. Columns 3-6 depict critical value of $a$, such that ASDE phase exists (i.e. ASDE exists if $a>a_{\rm cr}$). In all cases, a necessary requirement is that $k<3$. Columns 7-9 depict the asymptotic temporal slope obtained for jets with $a<a_{\rm cr}$ once the emission is dominated by increasing latitudes (sASDE phase).\\
    $^\dagger$For PLS E, the emission becomes dominated by the contribution from small radii for $k > 23/13$; 
    this regime is not yet fully explored even for a spherical flow.}
    \label{tbl:acr}
\end{table*}

Jets with very shallow angular energy structures are not favored by GRB observations (see \S \ref{sec:Intro}). Nonetheless, it is worth recalling that a structure with $a\sim 2$ and $b\sim 0$, while being somewhat fine-tuned, cannot at this stage be directly ruled out based purely on observational evidence.\footnote{Structure with $a\sim 2, b\sim 0$ result in afterglow lightcurves similar to those from steep structure jets observed close to their cores (e.g. \citealt{Granot-Kumar-03}), while a flat LF distribution is required also so that the observed correlation between early X-ray afterglow flux and prompt $\gamma$-ray fluence can be maintained \citep{Beniamini-Nakar-19}.} 
Even leaving such consideration aside, shallow jets may still contribute to some sub-population of GRBs. Finally, from a theoretical point of view, hydrodynamic simulations of lGRB jets that find such shallow structures (e.g. \citealt{Gottlieb2021}) provide good motivation to explore their observable features and find ways to test such models.

\begin{figure}
    \centering
    \includegraphics[width=0.48\textwidth]{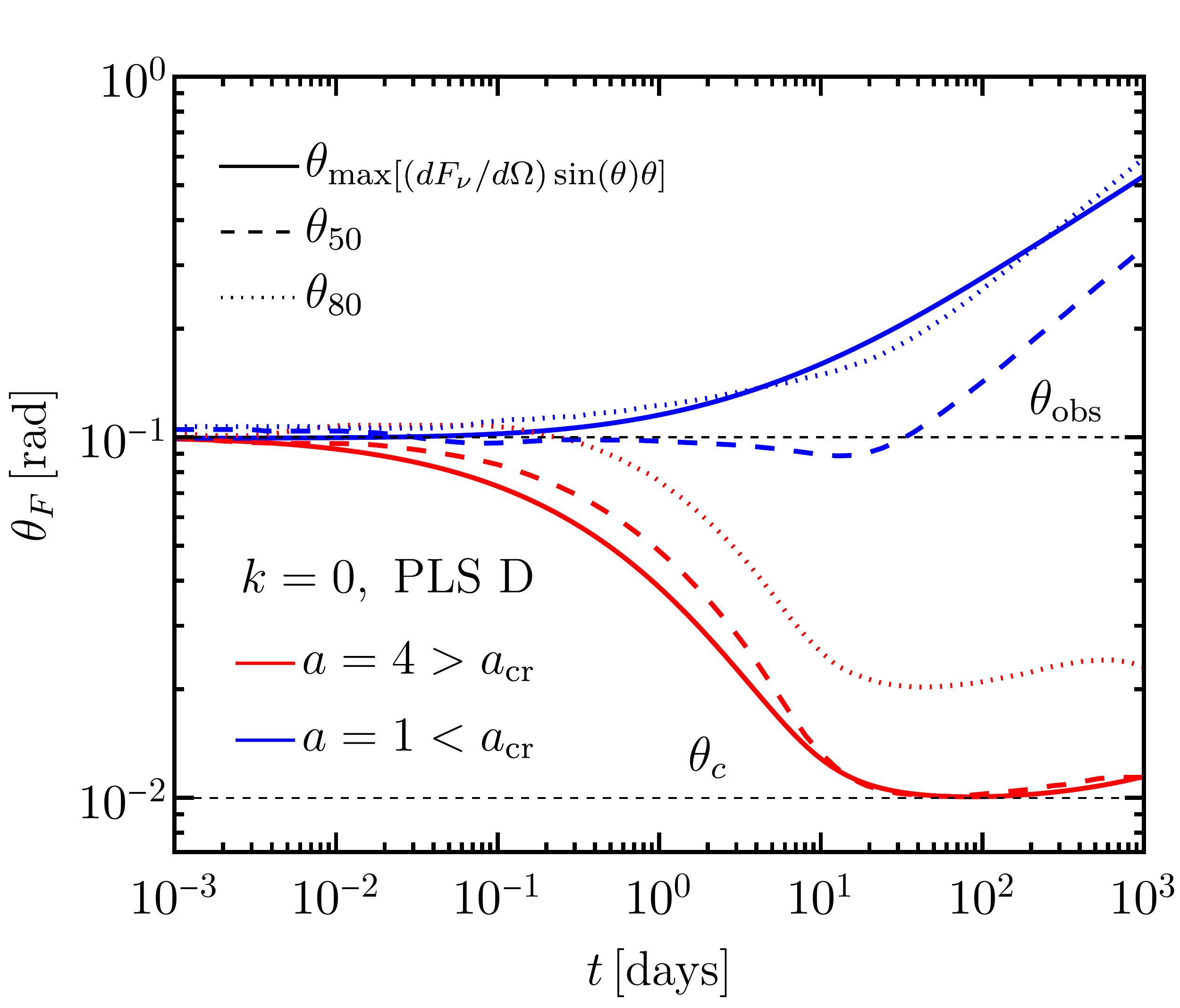}
    \caption{Temporal evolution of $\theta_F$ (polar angle from which emission dominates the observed flux) for different energy power-law 
    indices comparing steep ($a>a_{\rm cr}$) and shallow ($a<a_{\rm cr}$) jet angular profiles. The other model parameters, except for $\theta_{\rm obs}$, 
    are the same as assumed in Fig.\,\ref{fig:nu_m-nu_c-Vs-t}. The observed frequency here is $\nu=10^3\,$Hz that ensures that $\nu<\nu_m<\nu_c$ 
    (PLS D) over all time. Such a low $\nu$ is only chosen to clearly illustrate the difference between steep and shallow energy angular profiles. The 
    dashed and dotted lines indicate the angular extent of regions that contribute 50\% and 80\% of the total flux when integrated over solid angle 
    centered around the jet symmetry axis.}
    \label{fig:thf-diff-a}
\end{figure}

Jets with sufficiently shallow energy angular structures follow a unique temporal evolution. We define $a_{\rm cr}$ such that $a>a_{\rm cr}$ is required to obtain an ASDE phase. Recalling that during the ASDE phase $F_{\nu}^{\rm ASDE}\approx [\theta_{\rm min}\Gamma(\theta_{\rm min})]^2F_{\nu}^{\rm iso}(\theta_{\rm min})$ (where $F_{\nu}^{\rm iso}(\theta_{\rm min})$ is the flux density for an isotropic outflow with the same $\epsilon(\theta_{\rm min}),n,\epsilon_{\rm e},\epsilon_{\rm B},p,k$ and observed at the same time and frequency) and using $\epsilon\propto \theta_{\rm min}^{-a}$ during the ASDE we obtain $F_{\nu}^{\rm ASDE}(\theta_{\rm min})$. The condition that more inner material dominates at later times requires both that $d\log\theta_{\rm min}/d\log t<0$ (or equivalently $k<3$, i.e. that the flow decelerates rather than accelerates) and that $d\log F_{\nu}^{\rm ASDE}/d\log \theta<0$. This latter condition is used to obtain $a_{\rm cr}$for which $d\log F_{\nu}^{\rm ASDE}/d\log\theta=0$. The values of $a_{\rm cr}$ for the different synchrotron PLSs are given in Table \ref{tbl:acr}. For $k=0,p=2.2$, $1.3\lesssim a_{\rm cr}\lesssim 2.4$. For PLS D,E,G the value of $a_{\rm cr}$ tends to increase for larger $k$ (going up to $a_{\rm cr}=2.4$ for PLS D with $k=2$). This combined with the lower values of $a$ found in hydrodynamical simulations of lGRB jets (see \S\,\ref{sec:Jetstructure} and Table~\ref{tab:model-fit-params}) suggests that, particularly for lGRBs, it may be possible to probe the $a<a_{\rm cr}$ regime in at least part of the spectrum. Such a situation would provide a unique opportunity to probe different physical regimes at different points along the spectrum and in particular will be useful for placing strong constraints on the combination of $a,p,k$. As a simple example, if an ASDE phase is observed in PLS G but not in PLS H, then one can conclude that $k<4/3$ and $1.29\lesssim a \lesssim 1.53$.

\begin{figure*}
    \centering
    \includegraphics[width=0.95\textwidth]{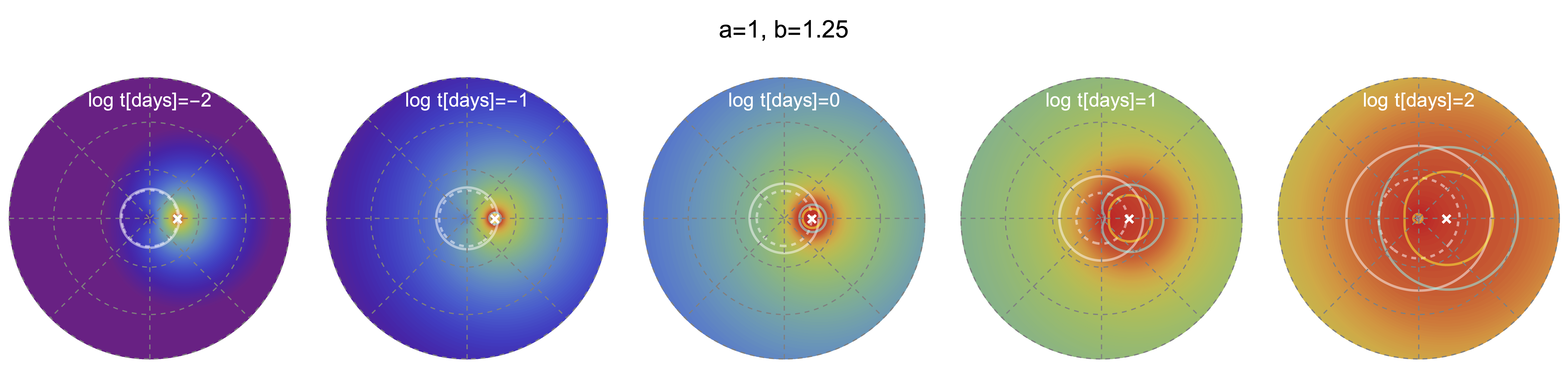}
    \includegraphics[width=0.95\textwidth]{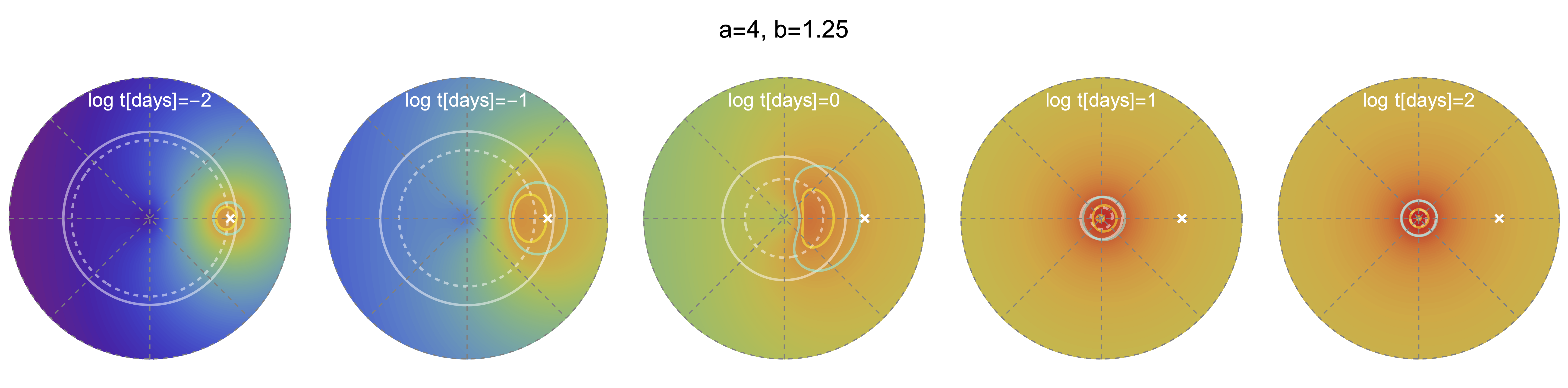}
    \centering\includegraphics[width=0.3\textwidth]{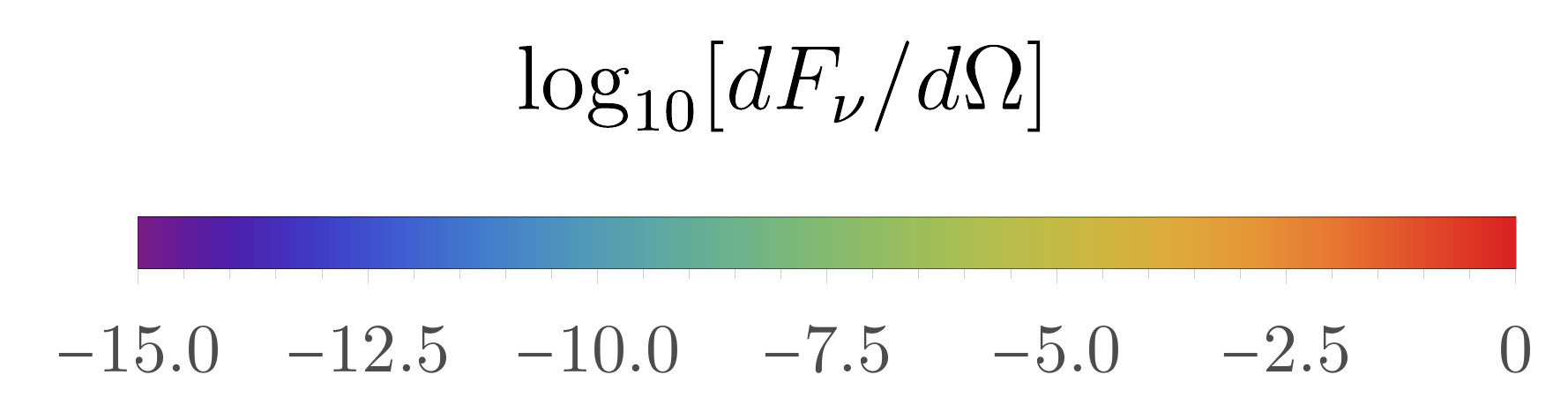}
    \caption{Angular map of $dF_\nu/d\Omega$ at different apparent time $t$ for two different energy per unit solid angle angular profiles, 
    with $a<a_{\rm cr}$ (top) and $a>a_{\rm cr}$ (bottom). Here $d\Omega$ denotes the solid angle centered around the jet symmetry axis. These do not represent the image of the outflow on the plane of the sky. The LOS of the observer is marked with a white cross at $\theta_{\rm obs}=0.1$ and the dashed gray concentric circles represent $\theta=0.174 (10^\circ)$, $\theta=0.35 (20^\circ)$, and $\theta=0.52 (30^\circ)$ angles. The bottom row shows a zoomed-in map that only extends to $\theta=10^\circ$. The yellow and cyan contours enclose the brightest angular regions contributing 50\% and 80\% of the total flux. The white solid and dashed contours enclose the angular regions that contribute 80\% and 50\%, respectively, of the total flux when integrated over $d\Omega$ from the jet symmetry axis.}
    \label{fig:flux-map}
\end{figure*}

We have shown above that for jets with $a<a_{\rm cr}$ material at angles smaller than the viewing angle is sub-dominant in its emission relative to that along the line of sight. This implies that such jets can eventually become dominated by material at angles {\it larger} than the viewing angle.
This is demonstrated in Fig.\,\ref{fig:thf-diff-a} that shows the temporal evolution of $\theta_F$ for a shallow ($a<a_{\rm cr}$) and steep ($a>a_{\rm cr}$) jet energy angular profile. The dominance of polar angles larger than $\theta_{\rm obs}$ at late times can be seen even more clearly in Fig.\,\ref{fig:flux-map}, that shows the angular map of $dF_\nu/d\Omega$ for shallow ($a<a_{\rm cr}$) and steep ($a>a_{\rm cr}$) jet structures. In a steep jet, the brightest 50\% and 80\% contributions to the flux, coming from angular regions marked with yellow and cyan contours respectively, moves to smaller polar angles (also shown in BGG20). Another way of seeing this behavior is by inspecting the integrated flux over the solid angle, $\int [dF_\nu(t,\theta,\phi)/d\Omega]d\Omega(\theta,\phi)$, centered at the jet symmetry axis, for which $\int[dF_\nu/d\Omega]\sin(\theta) d\theta\sim [dF_\nu/d\Omega]\theta^2$ (for $\theta\ll1$) is a proxy when considering axisymmetic flows. This is shown in the figure with solid and dashed white contours that enclose angular regions contributing 80\% and 50\% of the total flux. Again, for a shallow jet the maximum polar angle, up to which the integrated flux contributes a given fraction, always increases towards larger polar angles, whereas it always decreases towards smaller polar angles for steep jets.

In an analogous way to $\theta_{\rm min}$ defined in BGG20, we define an angle $\theta_{\rm max}$ which is appropriate for shallow jets:
\begin{equation}
\label{eq:thmax}
    \theta_{\rm max}-\theta_{\rm obs}\equiv \Gamma(\theta_{\rm max})^{-1}.
\end{equation}
This is the largest latitude from which emission is beamed towards the line of sight. At early times this angle is roughly constant $\theta_{\rm max}\approx \theta_{\rm obs}$. This continues until a critical time that we call $t_{\rm sh}$ that can be approximated by $\Gamma(\theta_{\rm obs},t_{\rm sh})=\theta_{\rm obs}^{-1}$ or 
\begin{equation}
 \tilde{t}_{\rm sh}\approx\xi_{\rm c}^{4-k \over 3-k} q^{8-2k-a \over 3-k}\approx \tilde{t}_{\rm pk} q^{-a\over 3-k}<\tilde{t}_{\rm pk}\ .
\end{equation}
Notice that $t_{\rm sh}\equiv t_{\rm beam}(\theta_{\rm obs})$, which is approximately the dip time for steep structure jets with double peaked lightcurves (see BGG20 for details). At $t\gg t_{\rm sh}$ we have $\theta_{\rm max}\approx \Gamma(\theta_{\rm max})^{-1}\approx \theta_{\rm obs} (t/t_{\rm sh})^{3-k \over 8-2k-a}$. This evolution leads to a distinct evolution of the lightcurves at these times, which we dub the ``shallow ASDE" or ``sASDE" for short. Analogously to the derivation  for steep jets above, we have $F_{\nu}^{\rm sASDE}\approx [\Gamma \theta_{\rm max}]^2F_{\nu}^{\rm iso}(\theta_{\rm max})\approx F_{\nu}^{\rm iso}(\theta_{\rm max})$. Plugging in the asymptotic temporal evolution of $\theta_{\rm max}$, this enables us to calculate the temporal slopes, $\alpha_{\rm s}$, of the different synchrotron PLSs. The results are given in Table \ref{tbl:acr}. These slopes are shallower than post-jet break slopes obtained for jets with $a>a_{\rm cr}$ (but steeper than the pre-jet break slopes for $0<a<a_{\rm cr}$). 

Putting together the asymptotic slopes and characteristic times, the lightcurve for shallow jets is given by
\begin{equation}
\label{eq:Fsh}
    F\!=\!F_{\rm sh} \bigg[1\!+\!\bigg(\frac{t}{t_{\rm dec}(\theta_{F,0})}\bigg)^{-2}\bigg]^{\alpha_{\rm d}\!-\!\alpha_{\rm i}\over 2} \bigg(\frac{t}{t_{\rm sh}}\bigg)^{\alpha_{\rm d}}\bigg[\frac{1\!+\!(t/t_{\rm sh})^2}{2}\bigg]^{\alpha_{\rm s}\!-\!\alpha_{\rm d}\over 2}
\end{equation}
where $F_{\rm sh}$ is the flux at $t_{\rm sh}$. To a first approximation, $F_{\rm sh}$ is the flux due to material near $\theta_{\rm F,0}$ as recorded by an observer on-axis to that material at a time $t_{\rm sh}$. The lightcurve approximated by Eq.\,(\ref{eq:Fsh}) is valid until $\theta_{\rm max}\approx 1$, after which (i) $\theta_{\rm max}$ can no longer continue to grow and (ii) the material dominating the observed emission is no longer ultra-relativistic. Beyond this point in time, the flux would decay as per a standard non-relativistic outflow.

\begin{figure}
    \centering
    \includegraphics[width=0.48\textwidth]{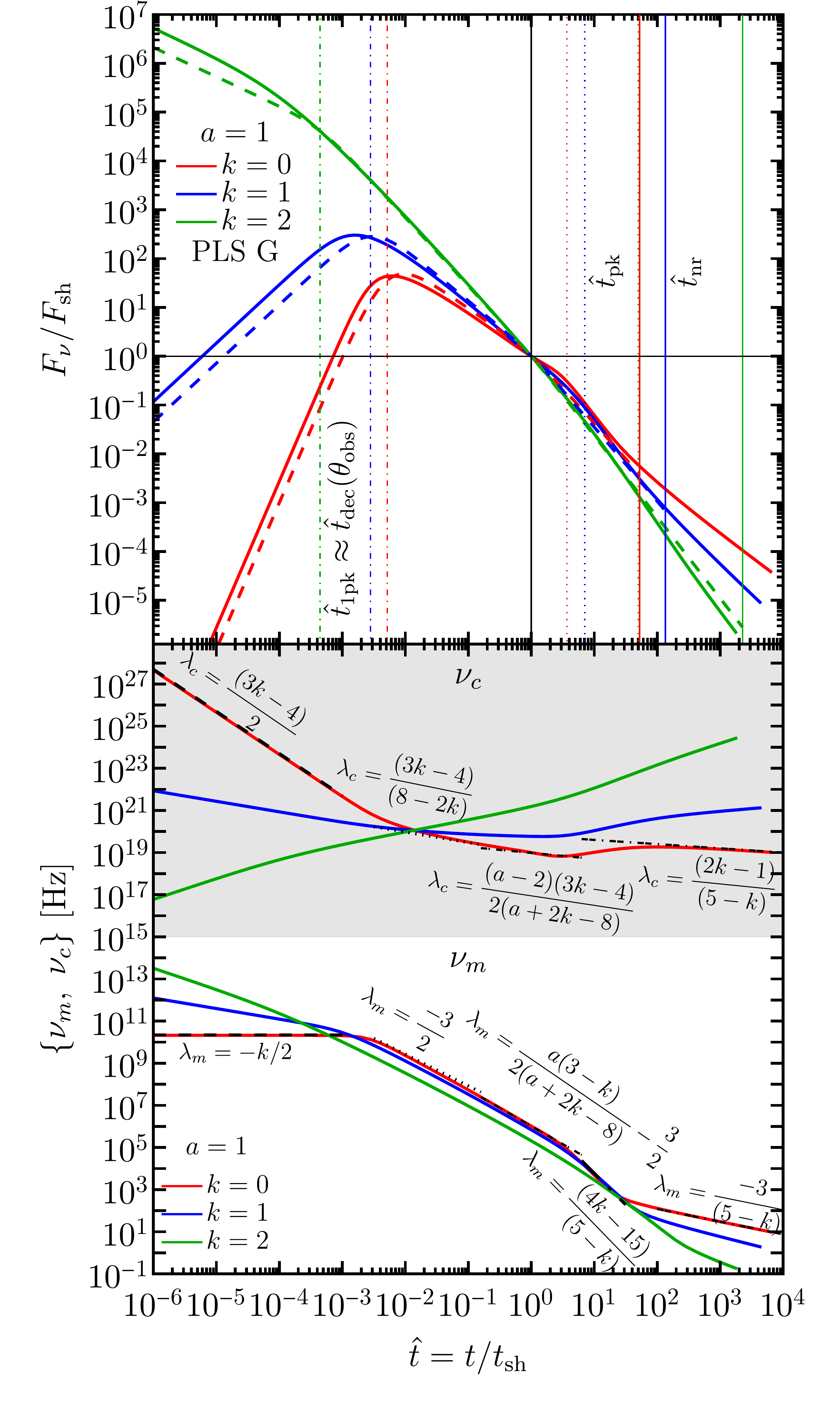}
    \caption{Lightcurves (top) and temporal evolution of synchrotron critical frequencies (bottom), shown for an outflow with shallow ($a<a_{\rm cr}$) 
    energy angular structure as well as for different external density radial profiles ($k$). The horizontal axis is apparent time $t$ normalized by 
    the time $t_{\rm sh}$ beyond which angles $\theta>\theta_{\rm obs}$ start dominating the observed flux. All timescales with a hat symbol are normalized 
    by this time. In the top-panel, the vertical axis shows flux density normalized by the flux $F_{\rm sh}=F_\nu(t_{\rm sh})$. The colored dashed lines are obtained 
    from the analytic approximation of the lightcurve as given in Eq.\,(\ref{eq:Fsh}), 
    and the solid lines are obtained from the full numerical integration. The different critical times are shown with a dash-dotted ($\hat t_{\rm 1pk}$), 
    dotted ($\hat t_{\rm pk}$), and thin solid ($\hat t_{\rm nr}$) lines. In the bottom-panel, the gray shaded region shows the curves for $\nu_c$ and the 
    unshaded region shows the same for $\nu_m$. The asymptotic temporal slopes in different segments are shown with different line traces along with their 
    corresponding analytic expressions. The assumed model parameters are the same as in Fig.\,\ref{fig:nu_m-nu_c-Vs-t}. Extreme values for $\nu_m$ and $\nu_c$ 
    are shown only to illustrate the full range of possible temporal evolution.}
    \label{fig:num-nuc-tobs-shallow-jets}
\end{figure}

Finally, we also use the evolution of $\theta_{\rm max}$ to calculate the evolution of $\nu_{\rm m},\nu_{\rm c}$ at $t>t_{\rm sh}$. 
The results are shown in Table\,\ref{tab:numnuc}, and complete temporal evolution along with the lightcurves is shown Fig.\,\ref{fig:num-nuc-tobs-shallow-jets}. 
As opposed to the ASDE phase, we see that here the evolution of the characteristic frequencies explicitly depends on the jet structure.

We conclude that shallow jets (with $a<a_{\rm cr}$) lead to a distinct evolution from steep structure lightcurves in terms of the critical break time, the flux evolution before and after the break and the evolution of the synchrotron frequencies. A final difference is that the flux centroid in such jets will progressively move away from the jet core, rather than towards it and will exhibit a more elongated shape.

\begin{figure}
    \centering
    \includegraphics[width=0.48\textwidth]{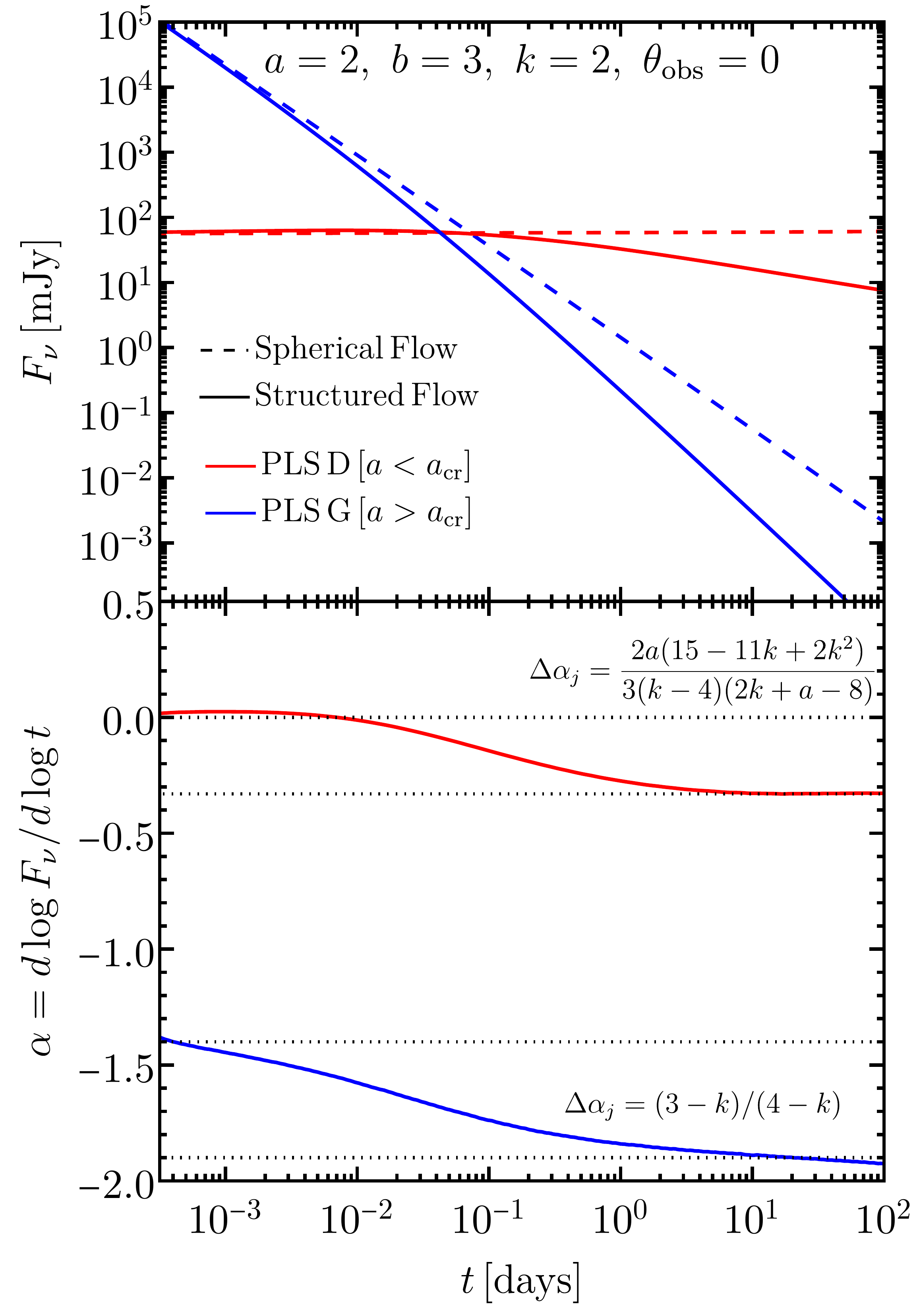}
    \caption{Chromatic jet breaks in an angular structured flow can arise for some values of $a$ that lie above and below $a_{\rm cr}$ for different PLSs. (Top) The lightcurves for an on-axis ($\theta_{\rm obs}=0$) observer from an angular structured flow and for different PLSs are compared with the same for a spherical flow. (Bottom) The change ($\Delta\alpha_j$) in the flux ($F_\nu\propto t^\alpha$) temporal power-law index shows chromatic behavior across different PLSs (see Table\,\ref{tbl:acr} for PLS definitions). The assumed model parameters are: $\theta_c=0.01$, $\epsilon_c=10^{53}\,{\rm erg\,str}^{-1}$, $\Gamma_c=2\times10^3$, $n(R_0=10^{18}\,\rm{cm})=10^{-2}\,{\rm cm}^{-3}$ ($A_\star=3.3\times10^{-2}$), $\epsilon_e=0.05$, $\epsilon_B=10^{-4}$, and $p=2.2$, with $\nu_{\rm PLS\,D}=10^5\,$Hz and $\nu_{\rm PLS\,G}=10^{17}\,$Hz. This set of fiducial parameters are chosen to gain enough dynamic range so as to clearly demonstrate the chromatic behavior.}
    \label{fig:chromatic-breaks}
\end{figure}

\subsection{Chromatic Jet Breaks in On-Axis Shallow Jets}

As remarked earlier, the difference in post jet-break temporal slopes for $a<a_{\rm cr}$ as compared to $a>a_{\rm cr}$ leads to interesting chromatic behavior. When considering on-axis observers ($\theta_{\rm obs}<\theta_c$) and when $a>a_{\rm cr}$ for a given PLS, the post jet-break temporal power-law index $\alpha_{\rm f}=d\log F_\nu/d\log t$ for a structured flow is the same as obtained for a top-hat jet, due to the fact that for apparent times after the jet break time, $t>t_j$, $\theta\sim\theta_c$ dominates the flux. Therefore, the change in the flux temporal slope is \textit{achromatic} and the same in both cases (and independent of $a$ for $a>a_{\rm cr}$), i.e. $\Delta\alpha_j=(3-k)/(4-k)$ for non-spreading jets.\footnote{Note that this is the asymptotic value of $\Delta\alpha_j$ across the jet break, while just after the jet break the lightcurve decay slope is steeper than the asymptotic post-jet break value thus leading to an overshoot of this value \citep{Granot07,DeColle+12b}, which is more pronounced  for low $k$-values as well as in numerical simulations, even when accounting for their large asymptotic $\Delta\alpha_j$ values compared to non-spreading jets. It has been briefly noted in earlier works that $\Delta\alpha_j$ is smaller for an $a=1$ power-law jet \citep{Granot-Kumar-03} and that for a ring-like or fan-like jet $\Delta\alpha_j$ is half of that for a top-hat jet when viewed from within the jet aperture \citep{Granot-05}.} However, when $a<a_{\rm cr}$ polar angles larger than $\theta_c$ start to dominate the observed flux and this expression for $\Delta\alpha_j$ no longer holds. As shown in Table\,\ref{tbl:acr}, the value of $a_{\rm cr}$ changes across different PLSs. As a result, when $a<a_{\rm cr}$ in a given PLS and at the same time $a>a_{\rm cr}$ in another PLS, \textit{chromatic} jet breaks should be observed. This behavior is demonstrated in Fig.\,\ref{fig:chromatic-breaks} where $\Delta\alpha_j$ (bottom-panel) across the jet break is different in two PLSs. The pre-jet-break value of $\alpha=\alpha_d$ (see Table\,4 of BGG20) and the post-jet-break temporal slope is $\alpha=\alpha_s$, as given in Table\,\ref{tbl:acr}. In this regime $\Delta\alpha_j$ depends on $a$, smoothly and monotonously varying between $\Delta\alpha_j(a_{\rm cr})=(3-k)/(4-k)$ and $\Delta\alpha_j(0)=0$.

As on-axis GRBs are routinely observed, it is worth commenting on the compatibility of existing GRB data with shallow jet structures. Unfortunately, such a comparison is complicated by the fact that the prediction of shallow jets is to lead to smoother jet breaks, in which the change in temporal slope is reduced compared to the steep jet case. As such, and taking into account the noisiness of GRB data, there will be an observational bias against identifying such breaks. Furthermore, shallow breaks may also be interpreted instead as originating from a steep jet with a larger value of $k$. Breaking this degeneracy in $\Delta \alpha_j$, would become easier for jets observed off-axis, for which the different observed temporal and spectral slopes can provide enough information to unambiguously solve for $a,k,p$ (see \S \ref{sec:discuss1}) and for which the motion of the flux centroid can provide the ``smoking gun" evidence in favor of shallow jets.
Nonetheless, there are observational hints from current observations which are worth mentioning.
GRB 130427A is a very bright low-redshift burst, which has been monitored for several years. Despite this very long base-line of observations, the afterglow is consistent with a single PL temporal decay \citep{DePasquale2016} up to at least $80$\,Ms after the trigger, much greater than the expected time of a jet break. Such a behaviour can be understood in the context of a shallow jet, in which the temporal slope change during the jet break can become arbitrarily small (depending on $a$).
There are several other bursts which show achromatic breaks with small values of $\Delta \alpha_j\sim 0.5$ or which show very smooth jet break transitions \citep{Wang2018,Lamb2021} (e.g. GRB 050801 and GRB 051109A). Indeed, in a work studying dozens of GRBs with optical and X-ray afterglow data, \cite{Liang2008} have not found a single burst demonstrating achromatic breaks that satisfy the predicted closure relations for a jet break from a top-hat (or steep jet structure) GRB.
That being said, we caution that the temporal data around the jet break is often sparse and the specific fitting of the temporal lightcurves for a given burst can differ between different studies. This again emphasizes the difficulty of determining a shallow or steep structure, based on existing observations. A dedicated systematic study, constraining the possible existence of shallow structure jets in the on-axis GRB population could be very helpful to advance our understanding of GRB jets.

\section{Robust characteristics of GRB jet structures}
\label{sec:Jetstructure}

\begin{figure*}
    \centering
    \includegraphics[width=0.48\textwidth]{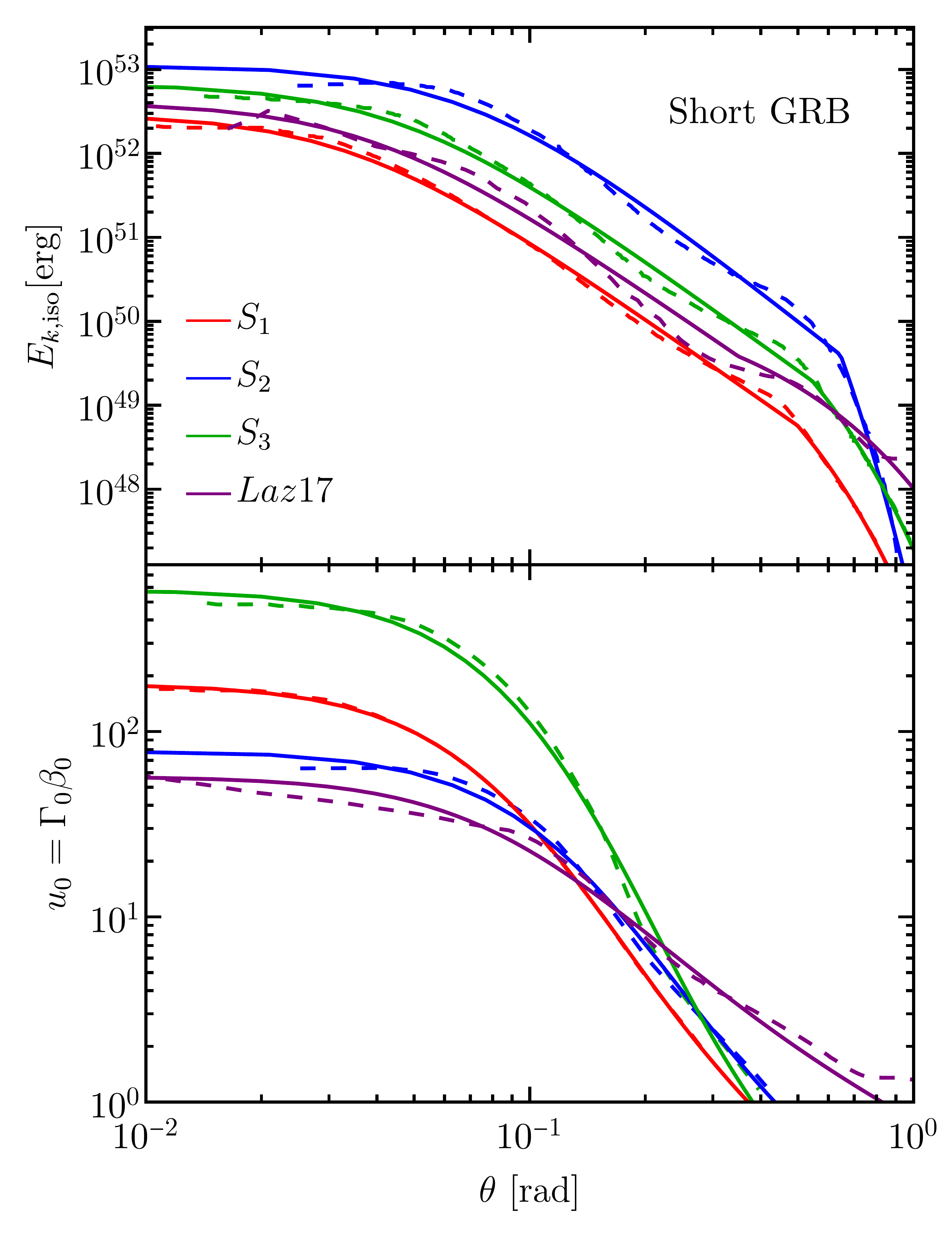}
    \includegraphics[width=0.48\textwidth]{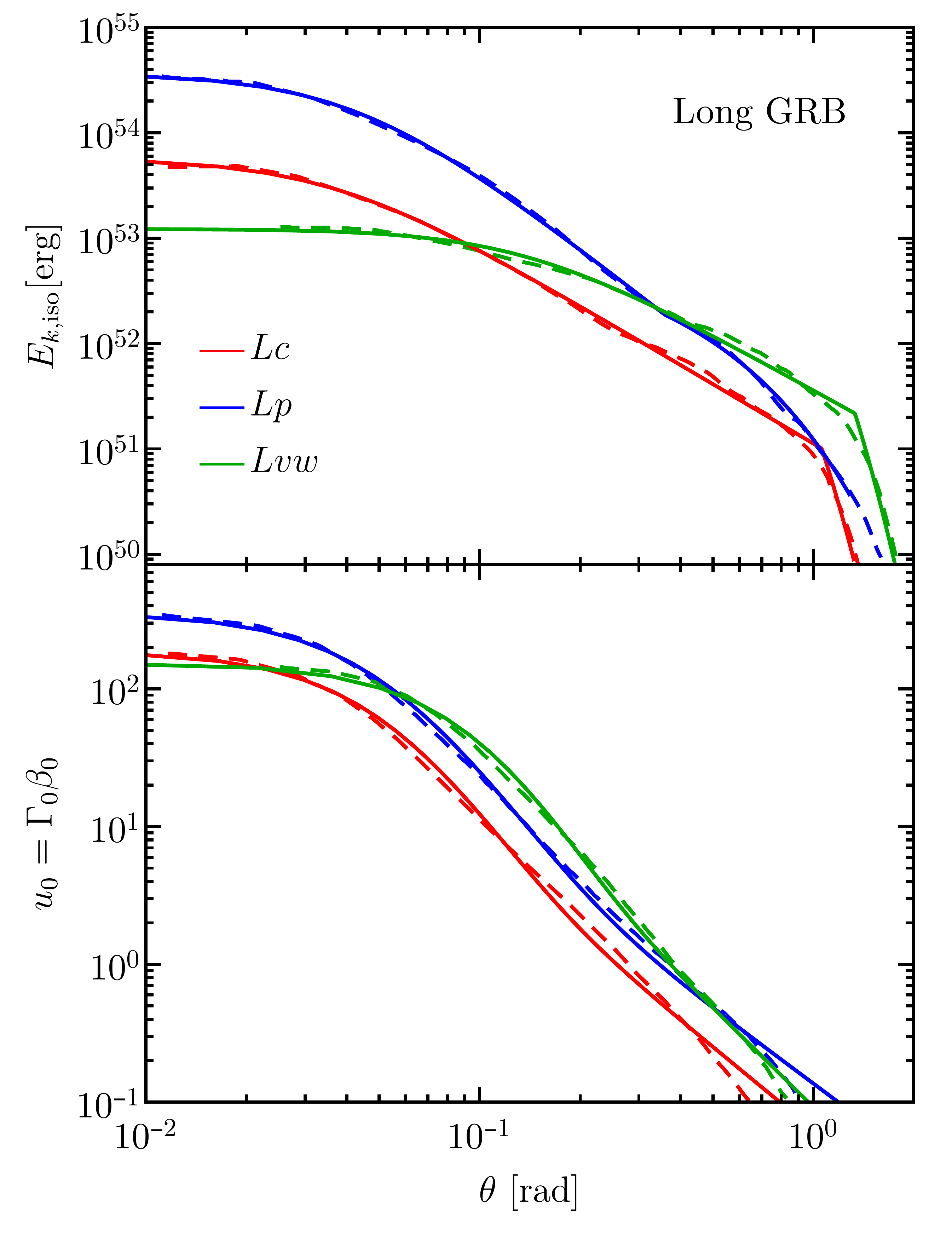}
    \caption{Angular distribution of the isotropic-equivalent kinetic energy $E_{\rm k,iso}(\theta)$ and initial proper four velocity $u_0(\theta)$ of the structured flow obtained for short-hard GRB (left) and long-soft GRB (right) jet models. Dashed lines are obtained from 3D hydrodynamic numerical simulations presented in \citet{Gottlieb2021} and \citet{Lazzati+17b}. Solid lines are model fits obtained from Eq.~(\ref{eq:struct-model}) and Eq.~(\ref{eq:PLJ}) for the energy and initial bulk-$\Gamma$ angular profiles, respectively.}
    \label{fig:model-profiles}
\end{figure*}

The post jet-breakout structure of GRB jets is governed by the properties of the incipient jet at the 
time of the launching close to the central engine, and by its interaction with the confining medium during 
propagation inside either the stellar interior (for long-soft GRBs) or dynamical ejecta (for short-hard 
GRBs). The physics of jet launching and the structure of the incipient jet are still very uncertain, however 
significant work, both analytical \citep[e.g.][]{Bromberg+11,Bromberg+14,Nakar-Piran-17,Lazzati-Perna-19,Hamidani+20,Hamidani-Ioka-21} 
and numerical \citep[e.g.][]{Ito+15,Duffell+15,Lazzati+17a,Lazzati+17b,Duffell+18,Harrison+18,Matsumoto-Masada-19,Gottlieb+20,Hamidani+20,Gottlieb2021,Hamidani-Ioka-21}, 
has been carried out to understand the jet propagation 
inside the confining medium. These works have helped in identifying some of the robust characteristics 
of GRB jets during propagation inside a dense medium as well as post-breakout. The properties of the 
incipient jet at launch time, although uncertain, depend on its magnetization, $\sigma=B'^2/4\pi h\rho'c^2$, 
which is the ratio of the magnetic field to matter proper enthalpy density and where $h$ is the enthalpy per unity rest-mass energy. In most hydrodynamic ($\sigma\ll1$) numerical simulations, the incipient jet is injected into a cone of half-opening angle $\theta_{j0}$ with radial velocity corresponding to a LF $\Gamma_{j0}$ such that $\Gamma_{j0}\theta_{j0}\simeq1$, and relativistically hot with $h_0=\Gamma_\infty/\Gamma_{j0}>\Gamma_{j0}\gg1$ (Some of these details may vary between such simulations done in different works). As the jet is barely in lateral causal contact it only slightly spreads laterally and largely retains $\theta_j\simeq\theta_{j0}$ before significantly interacting with the confining medium. Alternatively, the jet may be injected highly magnetized (with $\sigma_0\gg1$), only mildly relativistic ($\Gamma_{j0}\approx1$) and cold ($h_0\approx1$), 
such that\footnote{Note that some works define the magnetization parameter as $\tilde{\sigma}=B'^2/4\pi\rho'c^2=h\sigma$, in which case $\Gamma_\infty=(h_0+\tilde{\sigma}_0)\Gamma_{j0}$ where $\tilde{\sigma}_0=h_0\sigma_0$.} $\Gamma_\infty=(1+\sigma_0)h_0\Gamma_{j0}\sim\sigma_0$, and its properties at injection and evolution before breakout from the confining medium can be very different from the hydrodynamic case depending on the amount of magnetic dissipation, and as a result, on the evolution of $\sigma$ (see, e.g., discussion in \citealt{Bromberg+14}, and the simulation results of 
\citealt{Bromberg-Tchekhovskoy-16,Gottlieb+20,Gottlieb+22}).

The interaction of the relativistic jet with the confining medium slows it down and collimates it in a collimation shock, 
above which the jet becomes cylindrical. At the head of the jet, the jet material is shocked, slowed down, and channeled 
sideways, inflating a high-pressure inner cocoon. The jet-head typically moves at only mildly relativistic speeds, and 
drives a bow-shock like structure into the confining medium, which forms an outer cocoon, separated from the inner cocoon 
by a contact discontinuity. The jet continues to inject energy into the cocoon while making its way out of the confining medium. 
The drop in density at large distances typically causes the jet's head to accelerate. The contact discontinuity is 
Rayleigh-Taylor unstable as the \textit{lighter} (lower enthalpy density) shocked surrounding medium accelerates into 
the \textit{heavier} (higher enthalpy density) shocked jet material (i.e. $\ddot{r}_{cd}<0$ where $r_{cd}$ is the cylindrical 
radius of the contact discontinuity). The instability at the interface between the inner and outer cocoons grows as the 
jet is collimated by the cocoon. This leads to mixing between the two shocked fluids and, consequently, baryon-loading of 
the jet and a reduction of its local asymptotic LF, $\Gamma_\infty=\Gamma h$.

Upon breakout from the confining medium, both the jet and the cocoon continue to expand under their own 
pressure, both radially as well as laterally. Three-dimensional hydrodynamic simulations of relativistic 
jets breaking out of a homologously expanding dynamical ejecta consistently find three distinct components 
that constitute the angular structure of the axisymmetric post-breakout outflow. The jet core -- the 
ultrarelativistic and most energetic part of the outflow -- occupies the narrow angular region $\theta<\theta_c$ 
where it is characterized by an almost flat energy per unit solid angle $\epsilon(\theta)$, and a likewise 
flat proper velocity angular profile $u_0(\theta)$. At angles larger than the core angle ($\theta_c$), 
$\epsilon(\theta)$ declines as a power law with a characteristic power-law index that is set by the amount 
of mixing that occurred at the jet-cocoon interface while the jet was propagating inside the confining medium, 
as well as by the initial $\Gamma_\infty$ of the jet material as it is launched, which is usually approached 
at the core of jets that successfully break out of the confining medium. A similar power-law decline, albeit 
with a different power-law index, occurs in the $u_0(\theta)$ angular profile outside of the jet core. 

Angular distributions of both $E_{\rm k,iso}(\theta) = 4\pi\epsilon(\theta)$ and $u(\theta)$ obtained from the 3D 
hydrodynamic simulations of \cite{Gottlieb2021} for three different initial configurations 
of jets in short-hard (models $S_1$, $S_2$, $S_3$) and long-soft (models $Lc$, $Lp$, $Lvw$) GRBs are shown (with 
dashed lines) in Fig.~\ref{fig:model-profiles}. The power-law decline in $E_{\rm k,iso}(\theta)$ for both types 
of GRBs continues up to the point where the ejecta becomes non-relativistic. Beyond that, at angles 
$\theta>\theta_{\rm ccn}\simeq(0.35-1.3)\,$rad, the energy distribution shows a rapid exponential decline that results 
due to the angular structure of the cocoon that accompanies the relativistic jet.

\begin{table*}
    \centering
    \begin{tabular}{c @{\vline}| c|c|c|c|c|c | @{\vline} | c|c|c}
    \hline
          &\multicolumn{5}{c}{sGRB models} & & \multicolumn{3}{c}{lGRB models}\\ 
         & $S_1$ & $S_2$ & $S_3$ & th50 & gs50 & Laz17 & $Lc$ & $Lp$ & $Lvw$ \\
        \hline
        $\epsilon_{c,51}$ & $2.37$ & $9.06$ & $5.38$ & $1.86$ & $1.26$ & $40$ & $45.44$ & $290.46$ & $9.79$ \\
        $a$ & $3.22$ & $3.63$ & $3.33$ & $3.5$ & $2.82$ & $3.2$ & $1.88$ & $2.48$ & $1.78$ \\
        $\theta_{c,\epsilon}$ & $0.035$ & $0.072$ & $0.047$ & $0.043$ & $0.041$ & $0.04$ & $0.036$ & $0.043$ & $0.14$ \\
        $\theta_{\rm ccn}$ & $0.498$ & $0.645$ & $0.545$ & $\sim 0.42$ & $\sim 0.42$ & $0.35$ & $1.06$ & $0.36$ & $1.33$ \\
        $f_c$ & $10.67$ & $19.71$ & $10.25$ & - & - & $5.6$ & $9.39$ & $4.38$ & $7.78$ \\
        \hline
        $\Gamma_{\rm c,0}$ & $181.41$ & $79.07$ & $584.53$ & $58$ & $45.6$ & $57$ & $186.42$ & $353.29$ & $154.22$ \\
        $b$ & $4.02$ & $4.2$ & $5.76$ & $1.98$ & $1.62$& $2.1$ & $4.0$ & $3.61$ & $4.84$ \\
        $\theta_{c,\Gamma}$ & $0.084$ & $0.131$ & $0.113$ & $0.043$ & $0.041$ & $0.08$ & $0.057$ & $0.054$ & $0.115$ \\
        \hline
    \end{tabular}
    \caption{Parameters from the model fit to different jet structures. Fits to models: $S_1,S_2,S_3,Lc,Lp,Lvw$ of \citet{Gottlieb2021} are performed in this work. For comparison we show also fits to models th50 and gs50 by \citet{Nativi2021}, performed by \citet{Lamb2022} (due to the slightly different method of fitting the cocoon parameter, $f_c$, is not directly obtained from these fits) and a fit to the jet model presented in \citet{Lazzati+17b,Lazzati+18} which is performed by us and dubbed here Laz17. The fit 
    parameters obtained in this table are generally different from those obtained in the respective numerical studies due to the 
    slightly different angular profiles used here. In particular, the slope of the power-law wings of the velocity angular profile would change when fitting the 
    $u_0=\Gamma_0\beta_0$ angular profile, as done in \citet{Gottlieb2021}, as compared to the $\Gamma_0-1$ angular profile, as done here.}
    \label{tab:model-fit-params}
\end{table*}

We model the angular profiles from \citet{Gottlieb2021} using Eq.~(\ref{eq:PLJ}) and the best-fit model is 
shown (solid lines) in Fig.~\ref{fig:model-profiles}. To describe the angular structure of the cocoon, we supplement 
the formula for $\epsilon(\theta)$ in Eq.~(\ref{eq:PLJ}) with an exponential function for angles $\theta>\theta_{\rm ccn}$, 
such that
\begin{equation}\label{eq:struct-model}
    \frac{\epsilon(\theta)}{\epsilon_c} = \Theta_\epsilon^{-a}(\theta)\mathcal{H}(\theta_{\rm ccn}\!-\!\theta) 
    + \Theta_\epsilon^{-a}(\theta_{\rm ccn})e^{-f_c(\theta-\theta_{\rm ccn})}\mathcal{H}(\theta\!-\!\theta_{\rm ccn})\,,
\end{equation}
where $\mathcal H(\theta)$ is the Heaviside function. Such an exponential decline is not seen in $u(\theta)$, and 
thus for the $\Gamma_0(\theta)$ profile we use the power-law functional form from Eq.~(\ref{eq:PLJ}). Model fit 
parameters are presented in Table \ref{tab:model-fit-params}. When comparing the energy angular profiles obtained from 
the simulations, it is found that the short-hard GRBs generally show a steeper power law decline as compared to the long-soft
GRBs. Their proper four velocity angular profiles are, however, approximately similar. For both types of GRBs, the proper 
velocity profiles always show a steeper decline when compared with that of the energy, i.e. $b>a$, in the simulations 
from \citet{Gottlieb2021}.

These curves represent the initial angular structure of the flow post jet-breakout, which is then used to calculate the 
afterglow lightcurve under the assumption that there is no lateral spreading and each polar angle within the jet evolves 
as if it were part of a spherical flow with the local $\epsilon(\theta)$ and $\Gamma_0(\theta)$.

\section{Implications for short GRBs}
\label{sec:sGRBs}
\begin{figure}
\centering
\includegraphics[width = 0.4\textwidth]{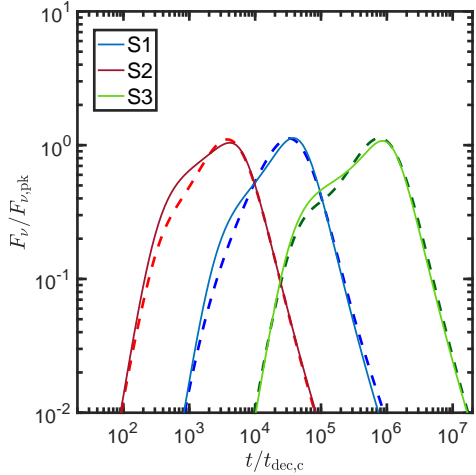}
\vspace{-1.4cm}
\caption{Lightcurves resulting from the three sGRB jet structures considered in the paper (see Table~\ref{tab:model-fit-params} for parameters), 
calculated for $\theta_{\rm obs}=0.35, k=0, p=2.16$. Time is depicted in units of the core deceleration time, and the flux is in units of the peak corresponding to the core of the jet coming into an off-axis observer's view. The shape of the lightcurves in these normalized units are independent of $n,\epsilon_e,\epsilon_B,d_{\rm L},\nu_{\rm obs}$. 
Dashed lines depict results from the analytical model described in \S \ref{sec:model} and BGG20, and solid lines depict results from direct 
numerical integration using the model of \citealt{GG2018}.}
\label{fig:S1S2S3fits}
\end{figure}

\cite{Gottlieb2021} focus on three sGRB structure models. To calculate the flux associated with the afterglows from those structures, one must further assume the (apriori unknown) values of the microphysical parameters $\epsilon_{\rm e}, \epsilon_{\rm B}$ and the external density $n$. As shown in BGG20, if we instead focus on the shape of the lightcurve, rather than the absolute flux, the lightcurve becomes largely independent\footnote{So long none of the characteristic synchrotron frequencies ($\nu_{\rm m},\nu_{\rm c},\mbox{ etc.}$) cross the observed band, and all the observations are done within a single PLS of the synchrotron spectrum.} of those unknowns and is uniquely described by the jet's structure and the viewing angle.
This is demonstrated in Figure \ref{fig:S1S2S3fits} showing the resulting lightcurves from our analytical model (see \S \ref{sec:model}) and comparing them with the full numerical integration over the emissivity from the structured jet. 

\begin{figure}
\centering
\includegraphics[width = 0.48\textwidth]{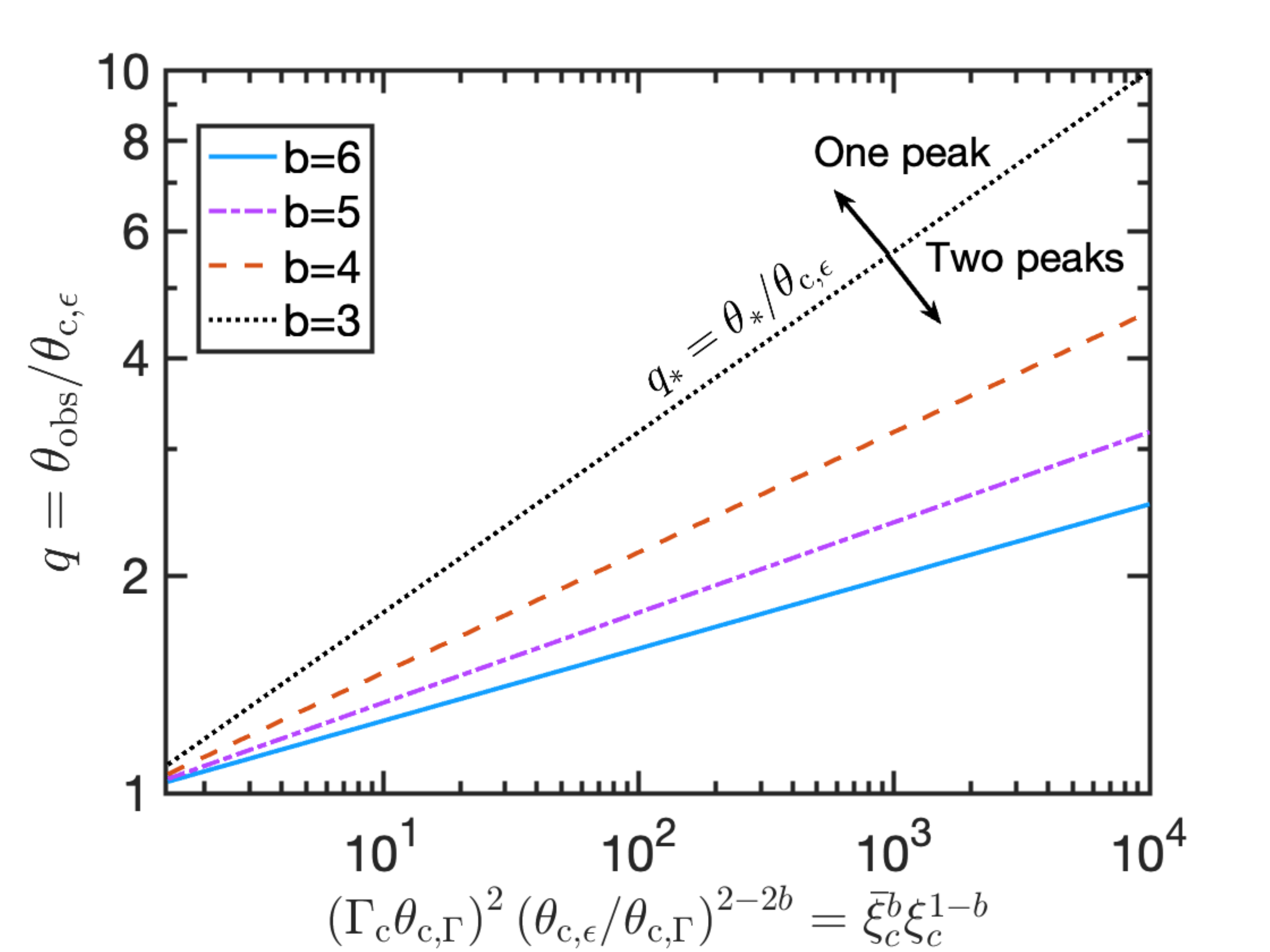}
\vspace{-0.4cm}
\caption{Type of afterglow lightcurve (single vs. double peaked) obtained for different GRB structures as constrained by numerical simulations. Each line represents the value of $q_*=\theta_{*}/\theta_{\rm c,\epsilon}$ for a given compactness of the core (specified by $b,\bar{\xi}_{\rm c},\xi_{\rm c}$, see \S\,\ref{sec:model} for details). $q>q_*$ (the space above each line) corresponds to single peak lightcurves and $q<q_*$ (the space below each line) corresponds to double peaked lightcurves. Short GRB jets are characterized by $4\lesssim b \lesssim 6$. Hydrodynamic long GRBs by $3.5\lesssim b \lesssim 5$.}
\label{fig:1vs2}
\end{figure}

As evident by Eq.\,(\ref{eq:thetastar}), the type of observed afterglow lightcurve (i.e. single or double peaked) depends mainly on three 
parameters: $b,q,\xi_{\rm c}$.\footnote{The energy structure only becomes important if $a\lesssim a_{\rm cr}$ (see \S\,\ref{sec:shallowjet} 
and Table~\ref{tbl:acr}), at which point the entire lightcurve evolution is modified.} The parameter $b$ as determined by simulations of 
sGRB jets and found to be $4\lesssim b \lesssim 6$ (see Table~\ref{tab:model-fit-params}). The allowed parameter range for one vs. two 
peaked lightcurves, calculated using Eq.~(\ref{eq:thetastar}), is depicted in Fig.\,\ref{fig:1vs2}. Two peak lightcurves will be more 
prevalent if sGRBs typically have highly relativistic cores ($\xi_{\rm c}\gg 1$)
and shallow Lorentz factor profiles (lower $b$). As a 
demonstration, $\theta_*=0.21, 0.27$, \& $0.27$ for $S_1$, $S_2$, and $S_3$ respectively.

\begin{figure}
\centering
\includegraphics[width = 0.45\textwidth]{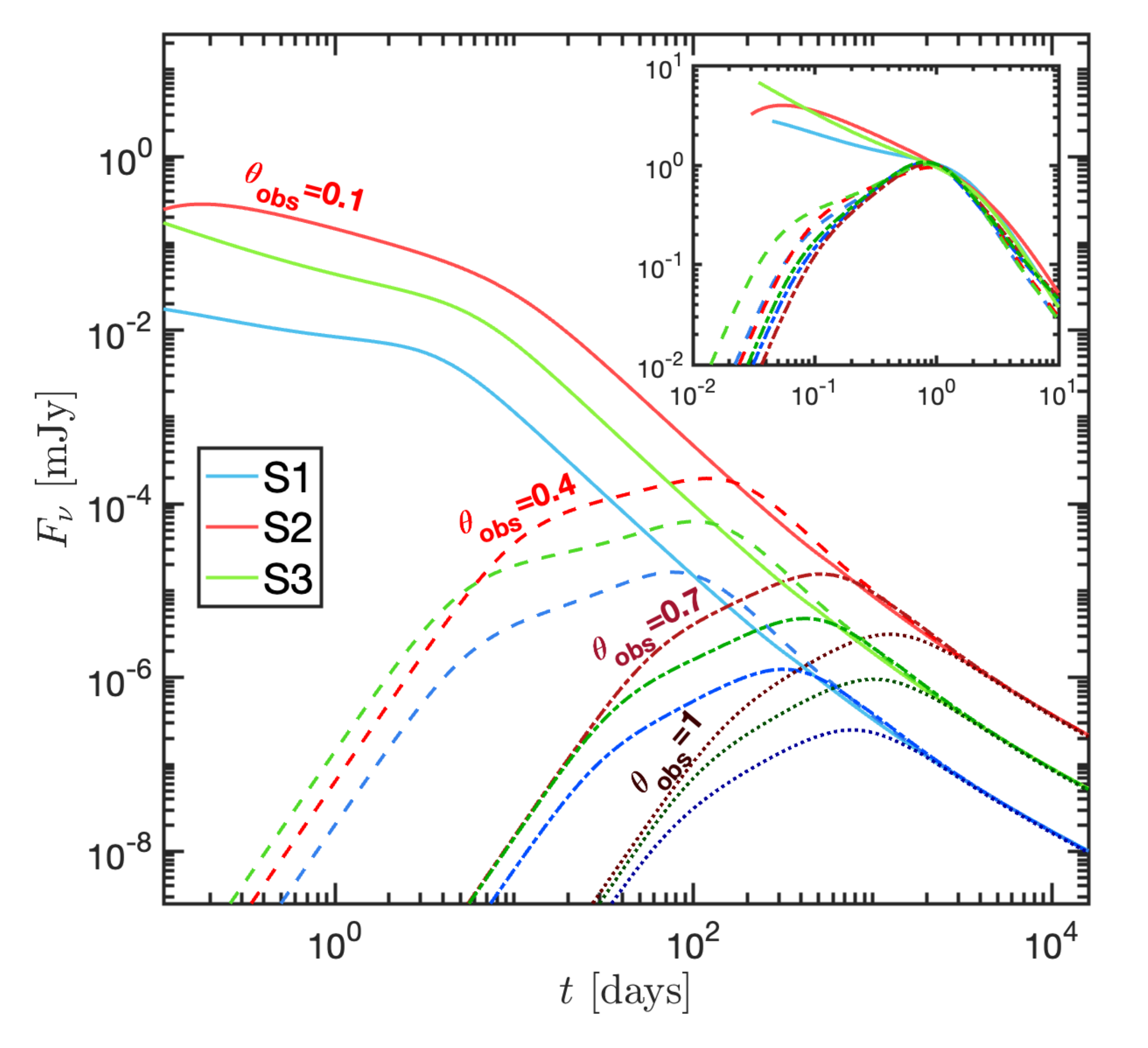}
\vspace{-0.2cm}
\caption{Lightcurves resulting from three sGRB models considered in this paper ($S_1$, $S_2$, $S_3$) for different viewing angles, $\theta_{\rm obs}=0.1,0.4,0.7,1$ (in solid, dashed, dot-dashed, and dotted lines respectively). Results in the main panel are plotted at a frequency of $10^{14}$\,Hz, as well as for $\epsilon_{\rm e}=0.01, \epsilon_{\rm B}=10^{-4}, n=0.01\mbox{ cm}^{-3}, p=2.16, d_{\rm L}=1.2\times10^{26}\mbox{ cm}$. An insert shows the same lightcurves (for the three smaller viewing angles) normalized such that the time is measured in units of $t_{\rm pk}$ and the flux in units of $F_{\rm pk}$. These lightcurve shapes are independent of $\epsilon_{\rm e}, \epsilon_{\rm B}, n, E_{\rm iso}, d_{\rm L}$.}
\label{fig:diffangles}
\end{figure}

While $S_1$, $S_2$, and $S_3$ can all reasonably explain the observed data from GW 170817 \citep{Gottlieb2021}, these models become distinguishable from each other once more off-axis sGRBs, mostly from lower viewing angles and observed at early times relative to their peak, are detected. For example, the transition from a double to a single peak for these models occurs at different $q$ values. The same is true for other measures of the lightcurve shape, such as the ratios of the fluxes and times of the two peaks (when there are two peaks), the ratio of times from the beginning of the shallow rise to the later peak and the slope of the shallow rise. This is demonstrated in Fig.\,\ref{fig:diffangles} where we show the lightcurves for all three models as a function of viewing angle.

\begin{figure}
    \centering
    \includegraphics[width=0.48\textwidth]{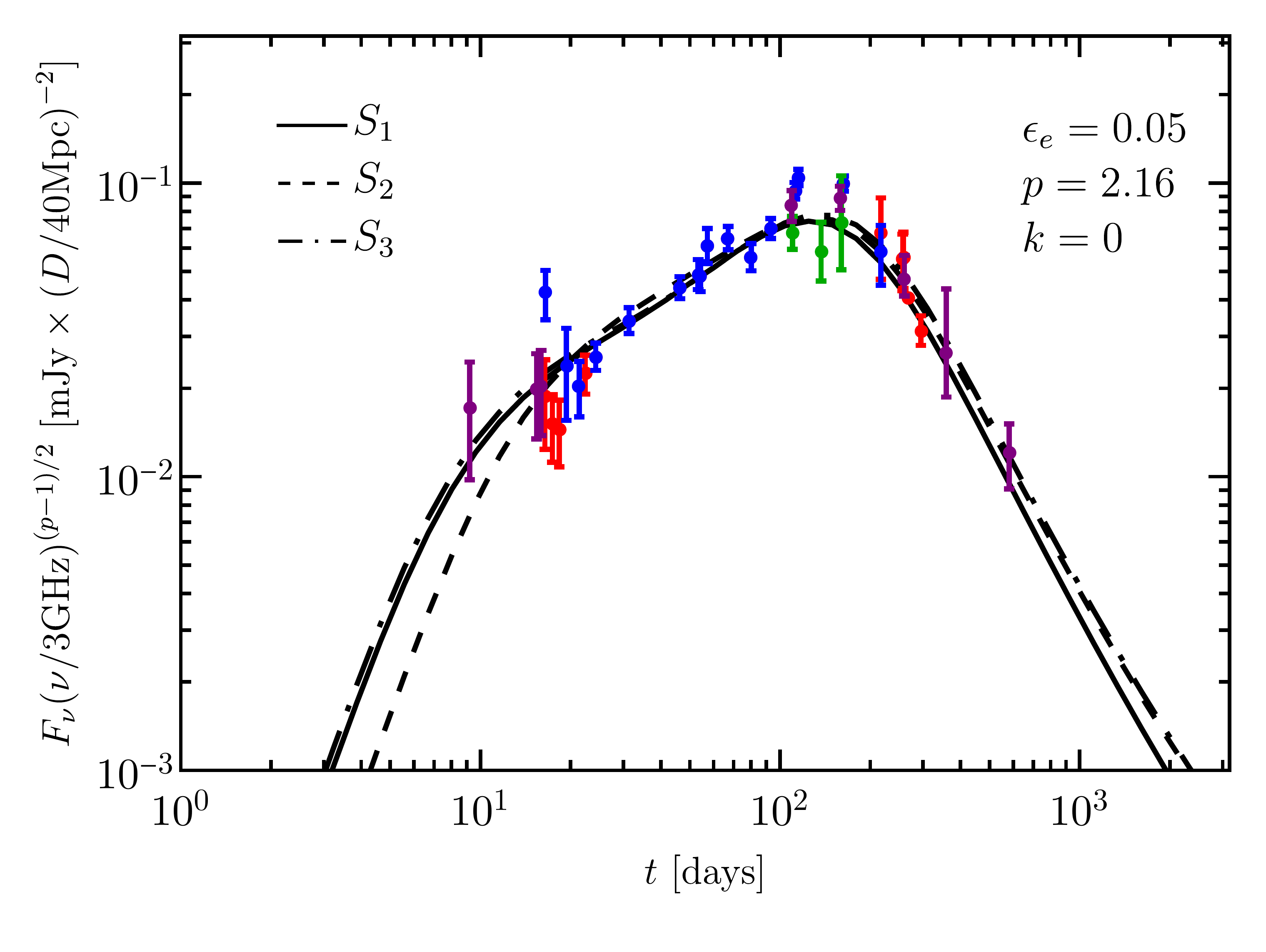}
    \vspace{-0.6cm}
    \caption{Model lightcurves from the sGRB angular structure profiles obtained from \citet{Gottlieb2021} compared with afterglow data of GW\,170817/GRB\,170817A. The model parameters adopted for the three lightcurves are: ($S_1$) $n = 8\times10^{-4}\,{\rm cm}^{-3}$, $\epsilon_B=6\times10^{-3}$, $\theta_{\rm obs}=0.35$; ($S_2$) $n = 2\times10^{-2}\,{\rm cm}^{-3}$, $\epsilon_B=4.5\times10^{-5}$, $\theta_{\rm obs}=0.45$; ($S_3$) $n = 10^{-2}\,{\rm cm}^{-3}$, $\epsilon_B=4\times10^{-4}$, $\theta_{\rm obs}=0.45$. These parameters are not unique and the model fits are degenerate. Only observations at times when the flow is still relativistic is shown here. At late times when the flow becomes non-relativistic lateral spreading becomes important, an effect not included in the afterglow code used in this work.}
    \label{fig:S-models-Vs-data}
\end{figure}

The comparison between the afterglow observations of GW\,170817 and model lightcurves derived from the jet angular structures profiles of \citet{Gottlieb2021} is shown in Fig.~\ref{fig:S-models-Vs-data}. Here we have extrapolated broad band observations to $3\,$GHz, which was possible owing to the fact that all the afterglow data was found to lie on a single power-law segment, corresponding to $p=2.15_{-0.02}^{+0.01}$, all the way up to $t\sim521-743\,$days post merger \citep{Hajela2019}. The lightcurves derived from the three different sGRB models fit the data well with some differences in the viewing angle, circumburst density, and shock microphysical parameters, all within a reasonable range. Only a rudimentary level of fitting to the data is done here and more careful fitting may find better agreement in the model parameters for the three angular profiles. However, the parameter space is degenerate \citep{Gill+19} and a unique set of model parameters cannot be obtained given the observations. In the figure, we only show afterglow data before the outflow starts to become non-relativistic. At later times, lateral spreading of the jet becomes very important and this affects the afterglow lightcurve \citep[e.g.,][]{Kumar-Granot-03,Rossi+04,Ryan+20,Lamb+21,Lamb+21b}. The afterglow code used in this work \citep[initially developed in][]{GG2018} does not account for lateral spreading and therefore it could not be used to fit the late time observations.

\section{Implications for long GRBs}
\label{sec:lgrbs}
For lGRBs, if the jets are hydrodynamic, the Rayleigh-Taylor instability leads to significant mixing of the jet with the pressurized cocoon that develops around it as it propagates. The result is jets with shallower energy and LF profiles with $2\lesssim a \lesssim 3$ and $3.5\lesssim b \lesssim 5$. The allowed parameter space for one vs. two peaks are shown in Fig.\,\ref{fig:1vs2}. The shallower LF profiles relative to jets suffering a smaller amount of mixing (such as in sGRBs) tends to increase the range of viewing angles (relative to the core) for which two peaks will be seen in the lightcurve. However, a more complete comparison must take into account also the change in $\theta_{\rm c,\epsilon}, \theta_{\rm c,\Gamma},\Gamma_{\rm c,0}$. Indeed, for the lGRB models explored in this work ($Lc$, $Lp$ and $Lvw$), we find that $\theta_{*}=0.12, 0.17,0.24$, respectively, is slightly lower than that found for the sGRB models in the previous section. Finally, we also note that the shallower energy structures (and potentially larger values of $k$, see below) of lGRBs, can lead to $a\lesssim a_{\rm cr}$, for which the lightcurve is no longer double-peaked and moreover is dominated by material close to the line-of-sight until late times (when it becomes dominated by $\theta>\theta_{\rm obs}$). As shown in \S\,\ref{sec:shallowjet}, such an evolution is clearly distinct from the single and double peaked lightcurves discussed in \S\,\ref{sec:model}.

\begin{figure}
\centering
\includegraphics[width = 0.45\textwidth]{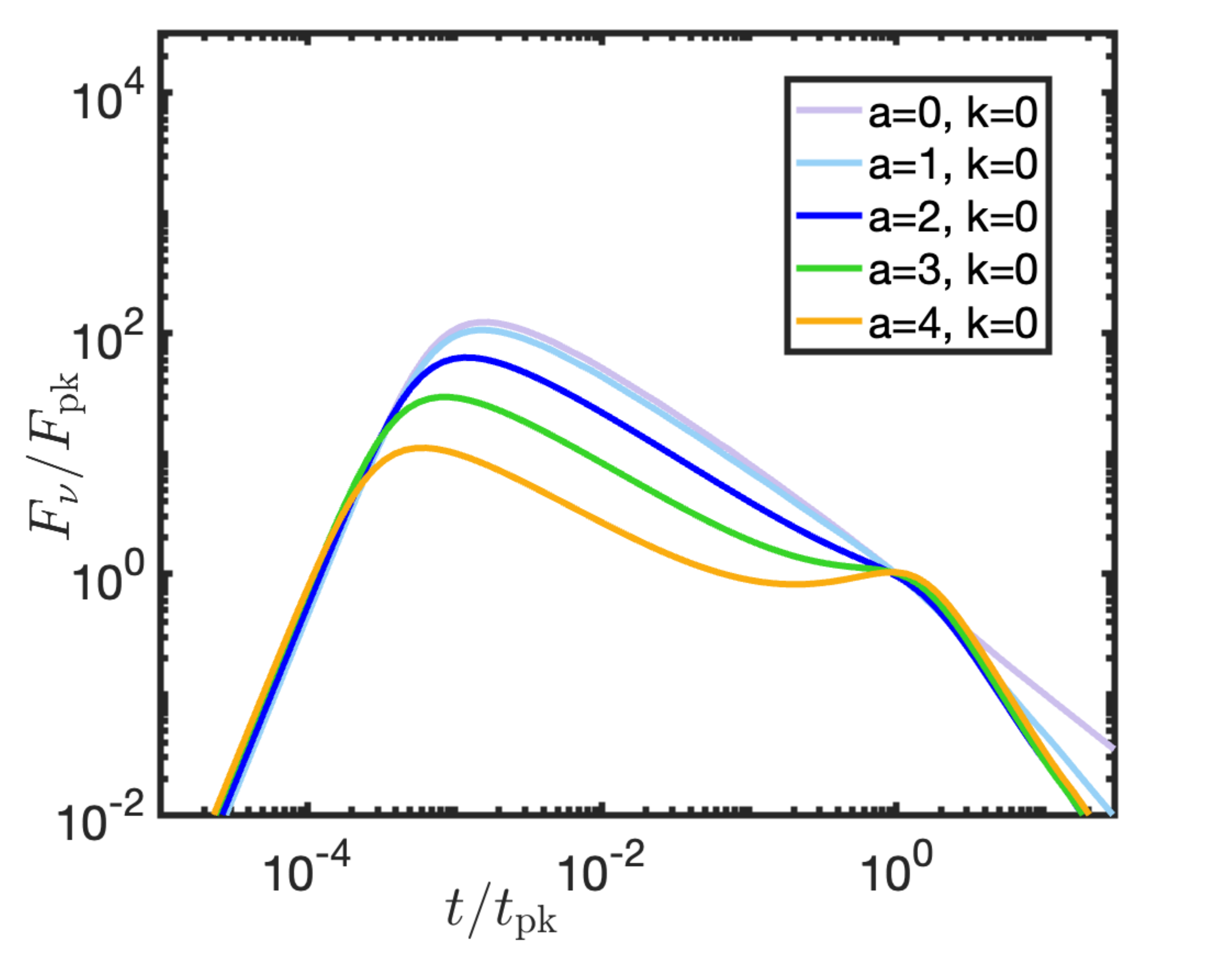}\\
\includegraphics[width = 0.45\textwidth]{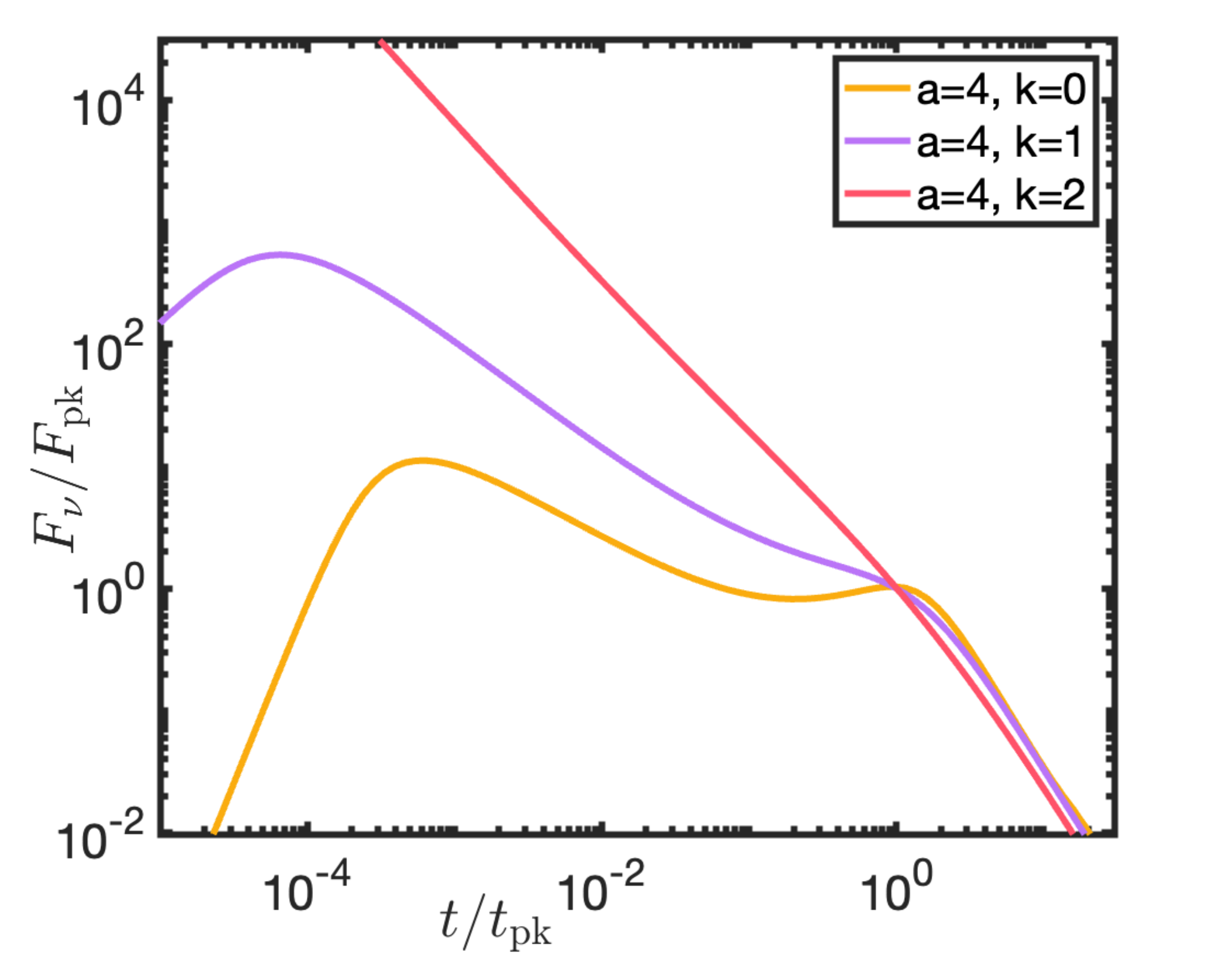}
\vspace{-0.2cm}
\caption{Lightcurves resulting from a GRB jet with $\xi_{\rm c}=500, q=4, b=1.5, p=2.16$ and different values of $a,k$. The observed frequency is such that the emission is in PLS G for the entire duration depicted. Time and flux are measured in units of those values at the time the core comes into view of an off-axis observer. For low $k$ and large $a$ this corresponds to the later (in case there are two peaks) afterglow peak. As $a$ decreases and\,/\,or $k$ increases, the two peaks are no longer distinct and the lightcurve becomes continuously declining after an early initial phase. For the lowest $a$ values, a shallow jet structure evolution is obtained.}
\label{fig:diffadiffk}
\end{figure}

The nature of the external medium can lead to notable differences between short and long GRBs. For lGRBs, the external density may be dominated by the wind ejected (at the rate $\dot M_w$ and with radial velocity $v_w$) from the progenitor star prior to its collapse. A progenitor stellar wind of constant $\dot{M}_w/v_w$ leads to $k=2$. However, detailed afterglow fitting of lGRBs suggests that in many cases, the lightcurves are better fitted with lower values of $k$ \citep[e.g.][]{Starling2008,VanderHorst2008} or even with $k=0$, expected for a uniform density \citep[e.g.][]{Panaitescu2002,Granot2005}. We therefore focus in what follows on a range of $k$ values, $k=0,1,2$. One notable feature is that for increasing $k$ and\,/\,or decreasing $a$, the temporal slope $\alpha$, related to emission dominated by progressively decreasing viewing angles (ASDE phase), decreases, and eventually becomes  negative. For such values of $\alpha$, the later peak (in situations where we would predict two peaks) is no longer a peak, and instead corresponds to a change in the temporal decline slope. This is demonstrated in Fig.~\ref{fig:diffadiffk} for the curves corresponding to $a=4,k=1$ or $a=3,k=0$. More generally, the two peaks (when this regime is encountered) are more clearly pronounced for greater $a$ and\,/\,or lower $k$. This can be seen by considering the difference in temporal slopes between the initial phase of the second peak and the later phase of the first: $\Delta \alpha = \alpha-\alpha_d$. The general expression for $\Delta \alpha$ is cumbersome, but simplifies significantly for particular $k$ values. We find
\begin{eqnarray}
\label{eq:Deltaalpha}
\Delta \alpha=  \left\{ \begin{array}{ll}\frac{3}{4}(3+p-\frac{8}{a}) & k=0 ,\\ \\
\frac{p}{2}+\frac{7}{6}-\frac{4}{a} & k=1 ,
\\ \\
\frac{p}{4}+\frac{1}{4}-\frac{2}{a} & k=2\ ,
\end{array} \right.
\end{eqnarray}
The increasing (decreasing) trend of $\Delta \alpha$ with $a$ ($k$) can be seen by comparing the $[k=0,a=2],[k=0,a=3],[k=0,a=4]$ ($[k=0,a=4],[k=1,a=4]$) 
cases in Fig.~\ref{fig:diffadiffk}. Finally, the shallow jet structure evolution discussed in \S\,\ref{sec:shallowjet} can clearly be seen in the figure for the cases $[k=0,a=1],[k=0,a=0]$. In particular, note the change in asymptotic slope compared to the steep jet lightcurves.

\begin{figure}
\centering
\includegraphics[width = 0.45\textwidth]{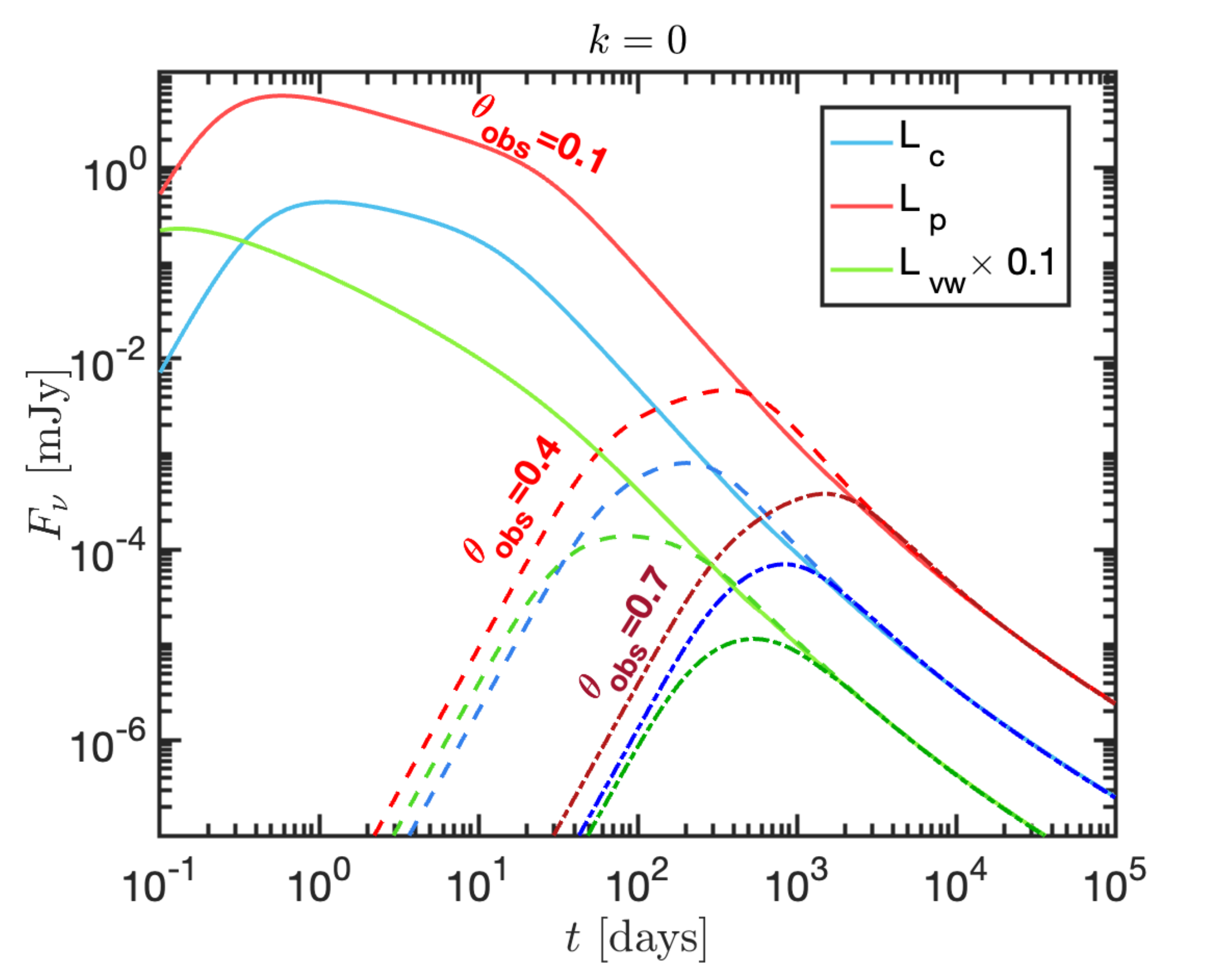}\\
\includegraphics[width = 0.45\textwidth]{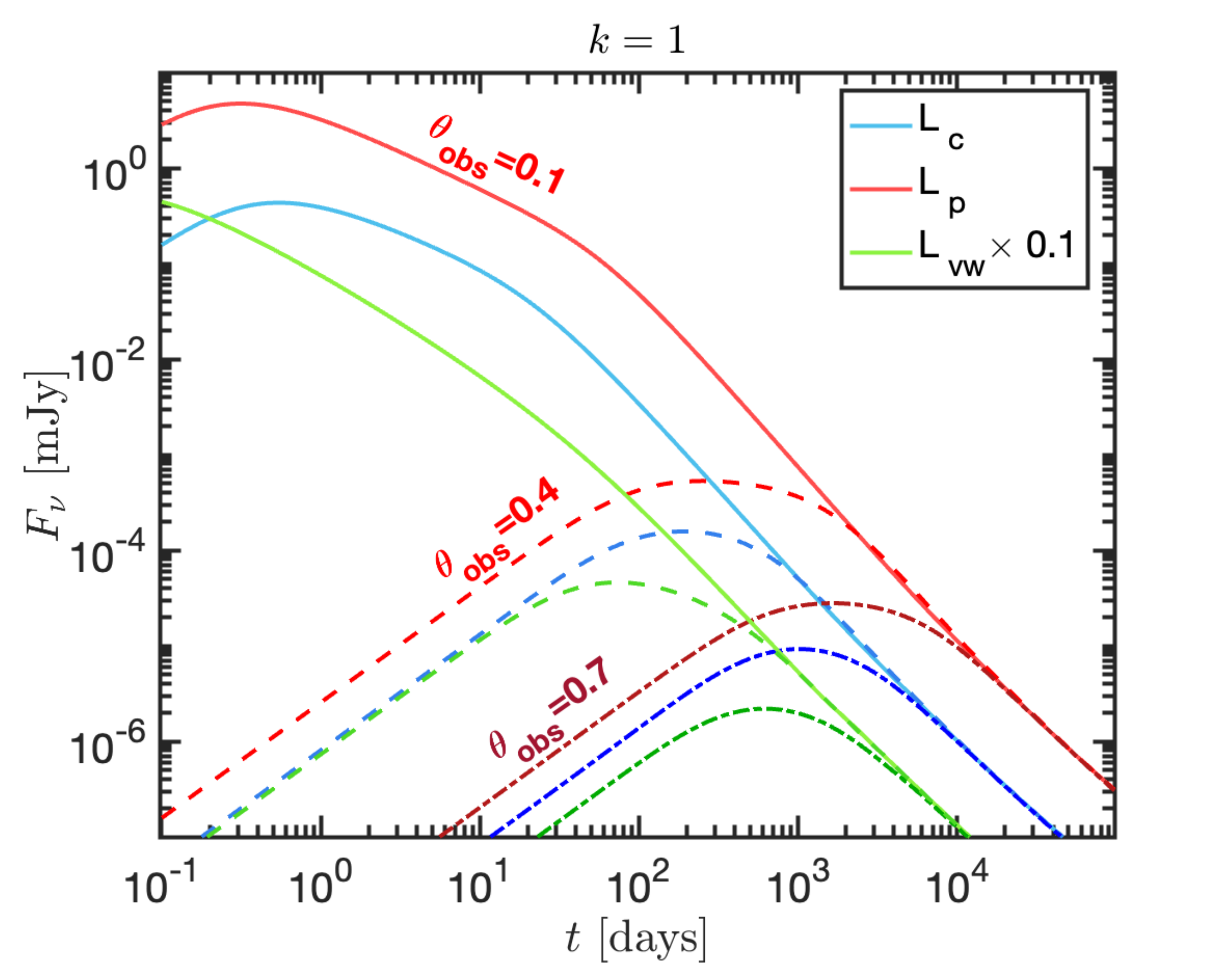}\\
\includegraphics[width = 0.45\textwidth]{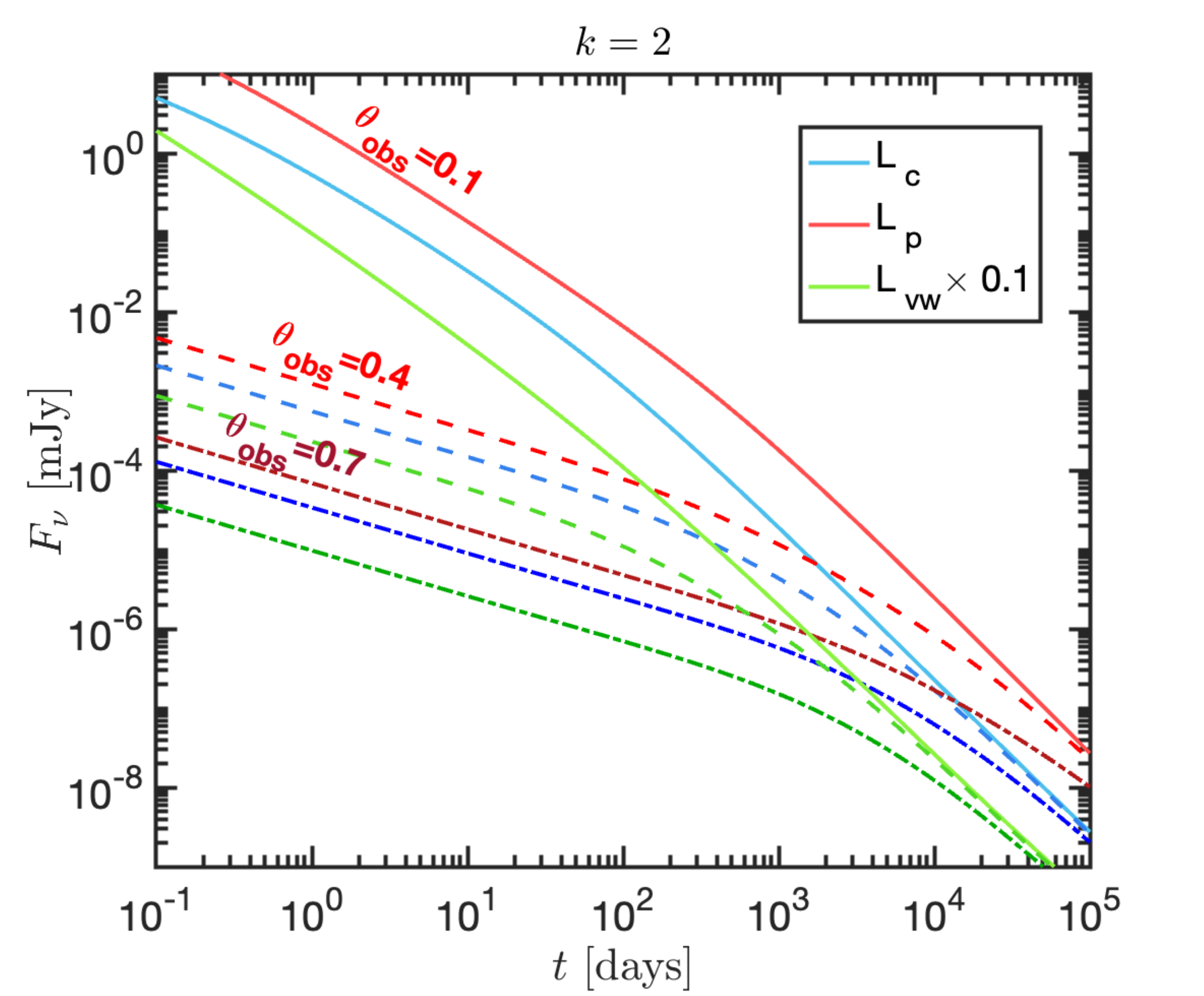}
\caption{Lightcurves resulting from three long GRB models considered in this paper ($L_{\rm c}, L_{\rm p}, L_{\rm vw}$) for different viewing angles, $\theta_{\rm obs}=0.1,0.4,0.7$ (in solid, dashed and dot-dashed lines respectively). Different panels show different values of $k$. Results are plotted at a frequency of $10^{14}$\,Hz, as well as for $\epsilon_{\rm e}=0.001, \epsilon_{\rm B}=10^{-4}, n(R_0)=0.01\mbox{ cm}^{-3}, R_0=10^{18}\,{\rm cm}, p=2.16, d_{\rm L}=1.2\times10^{26}\mbox{ cm}$ (note that the lightcurve shapes are independent of $\epsilon_{\rm e}, \epsilon_{\rm B}, n, E_{\rm iso}, d_{\rm L}$). }
\label{fig:lgrbcurves}
\end{figure}

Examples of afterglow lightcurves obtained for different structures, different viewing angles and different external media profiles (i.e. different $k$) 
are shown in Fig.~\ref{fig:lgrbcurves}. As for the sGRB structures, we find that different structures are most readily separated for lower viewing angles and for earlier observation times. As $k$ increases the later peak is delayed, and the lightcurve is dominated by the first two: $\alpha_{\rm i}, \alpha_{\rm d}$ or three (adding also $\alpha$) temporal slopes until very late times.

\begin{figure}
    \centering
    \includegraphics[width=0.48\textwidth]{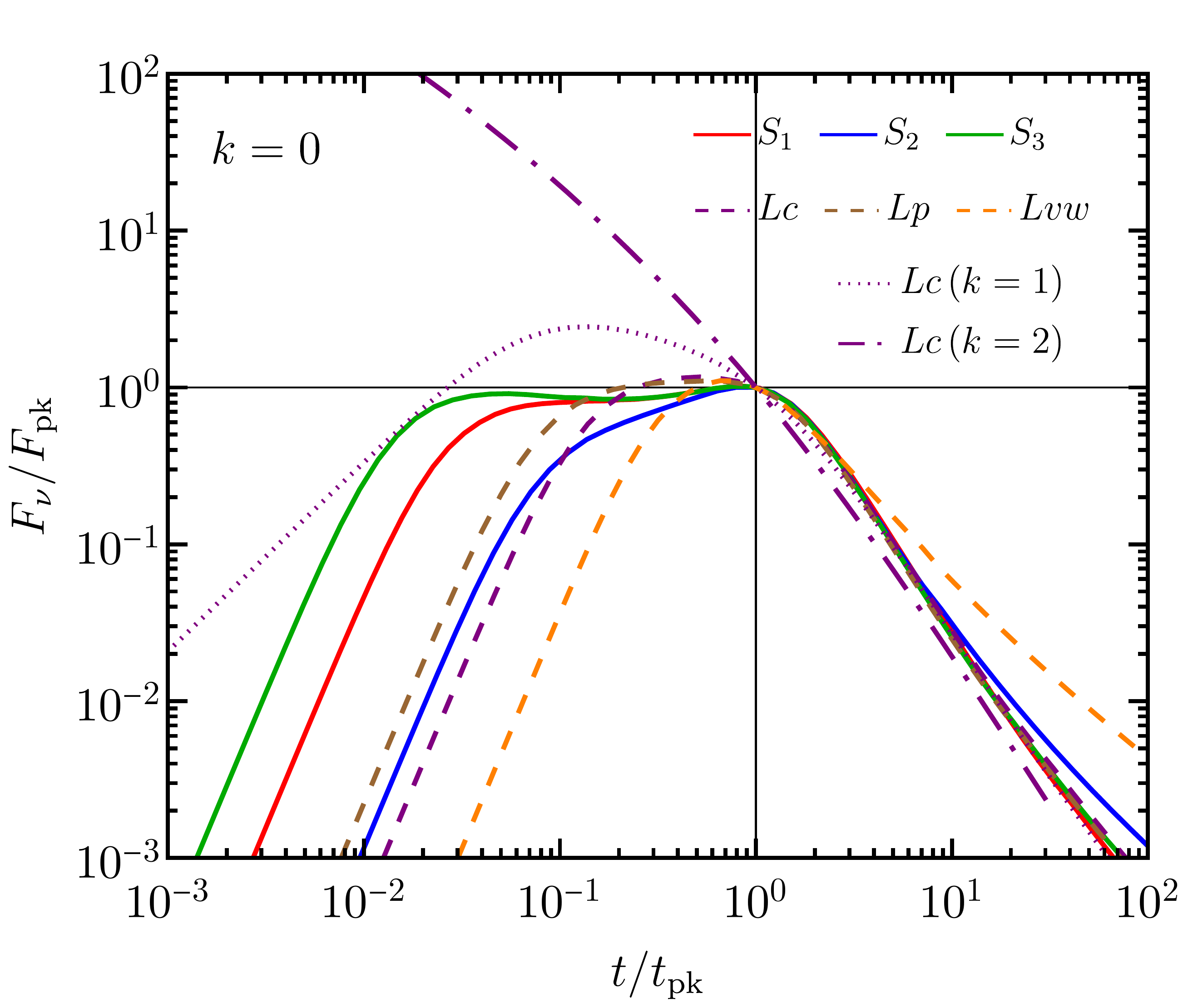}
    \caption{Comparison of lightcurves obtained for the short (solid) and long (dashed) duration GRB angular structure profiles obtained from the numerical hydrodynamical models of \citet{Gottlieb2021}. The time and flux density are normalized by the second peak time ($t_{\rm pk}$, or equivalently $t_{\rm sh}$ for the $Lvw$ model which has $a<a_{\rm cr}$) and the flux at that time ($F_{\rm pk}$, or $F_{\rm sh}$). The model parameters assumed here are: $\theta_{\rm obs}=5\theta_{c,\epsilon}$, $n=10^{-2},{\rm cm}^{-3}$, $\epsilon_e=10^{-2}$, $\epsilon_B=10^{-4}$, $p=2.16$. All the shown lightcurves obey $\nu_m < \nu < \nu_c$ (PLS G) and therefore are independent of $n$, $\epsilon_e$, $\epsilon_B$ when shown using normalized units. To compare with the $k=0$ case, example lightcurves for $k=1,\,2$ (valid for lGRBs) are also shown for model $Lc$.}
    \label{fig:sGRB-lGRB-Compare}
\end{figure}

We compare the lightcurves obtained for the different sGRB and lGRB angular structures profiles obtained from \citet{Gottlieb2021} in Fig.~\ref{fig:sGRB-lGRB-Compare} for a fixed set of model parameters. Significant differences can be seen between the model lightcurves at early times even for the same value of $k$, where the sGRBs show a hint of two peaks with a dip and the lGRBs show a rather smooth and broad hump (due to the lower values of $\theta_*$ as detailed above). In particular, we note the lightcurve for $Lvw$ that decays more shallowly than the other lightcurves, as in that case $a<a_{\rm cr}$ leading to an emission dominated by ever increasing latitudes at late times. In the figure, we again emphasize the different lightcurve behavior obtained for the lGRBs when $k$, the circumburst density profile power-law index, assumes different values. When compared with observations, such differences can be used to constrain the properties of the circumburst medium.

\section{Discussion}
\label{sec:discuss}
Three-dimensional hydrodynamic simulations of relativistic jets propagating through static (in lGRBs) and expanding (in sGRBs) media are rapidly improving our understanding of the angular structure that emerges due to the interaction of the jet with the circumburst medium. Some robust features, that include the (almost) flat jet core, power-law wings, and an exponentially declining cocoon, have emerged in the angular structure of both types of GRBs. In this work, we have derived parameterized model profiles using the results of the numerical simulations from \citet{Gottlieb2021} and tried to model the afterglow emission for a broad set of model parameters. We have used the detailed formalism developed in BGG20 that describes the temporal evolution of the afterglow emission from angular structured flows and extended it in this work in several ways, including also flows in which the core angles of the bulk $\Gamma$ and energy per unit solid angle profiles are not necessarily the same, or flows with very shallow angular structure. BGG20 have found that qualitatively different types of afterglow shapes (e.g. single or double peaked) are possible from the same underlying jet structure, depending on the viewing angle. The {\it shape} of the lightcurve (rather than the absolute normalizations of time and flux) depends only on a small number of parameters: the compactness of the core ($\xi_{\rm c}$) \footnote{In the case where the Lorentz factor core is different than the energy core, an additional parameter, quantifying the ratio of the two, is needed.}, the slopes of the energy and Lorentz factor angular profiles ($a,b$), the viewing angle relative to the core ($q$), the power-law slope of the accelerated electrons' energy distribution ($p$) and the slope of the external density radial profile ($k$). This is a significant simplification, as it gets rid of several parameters that are needed to describe the normalized fluxes and times (e.g. density, blast-wave energy and fraction of shock energy that is deposited in accelerated electrons and in magnetic fields), some of which are poorly constrained by the data. Despite this simplification, the number of independent constraints from a given observed off-axis jet, may still be insufficient to fully solve for all the parameters mentioned above, especially if some segments of the emission (for example, very early time data) are missing.

In the present work, we have included the temporal evolution of the characteristic synchrotron frequencies in the different phases of the afterglow when the outflow is misaligned ($\theta_{\rm obs}>\theta_c$). We find a unique phase of evolution for the characteristic frequencies that arises when the emission becomes continuously dominated by gradually decreasing latitudes (or polar angles $\theta$ from the jet symmetry axis) between the line of sight to the emitting material and the jet's core (the ASDE phase). At these times, the temporal evolution of the frequencies depends only on the radial profile of the external density (characterized by $k$) while the normalization carries information about the properties of the jet structure along the line of sight to the observer. Therefore, measuring the rate of evolution of these frequencies could provide an important and independent constraint on the structure parameters and the viewing angle which could help remove the degeneracies discussed above.

As we demonstrate in this work, the temporal evolution of the characteristic synchrotron frequencies enables the determination of the viewing geometry and circumburst medium radial profile in off-axis events. In typical cosmological GRBs, we observe emission from the decelerating core as the jet is viewed on axis, and for which the temporal evolution of $\nu_m$ and $\nu_c$ is much simpler and well understood \citep{Sari+98,Granot-Sari-02}. In the off-axis jet case, tracking the evolution of $\nu_m$ and $\nu_c$ can be challenging observationally as it spans a broad range in frequency over a broad range in time. In some cases, like GW\,170817, it might not even be possible at all, since in this particular case there were no definitive signs of either $\nu_m$ or $\nu_c$ crossing and the entire broad band observations only sampled PLS G of the synchrotron spectrum despite the event being observable for several years in a wide frequency range, from radio up to X-rays. 

\subsection{What model parameters can be constrained from the afterglows of misaligned structured jets?}
\label{sec:discuss1}
As explored by BGG20, the underlying GRB properties can be constrained by the shape of the observed lightcurve. The temporal slopes during the different phases of emission ($\alpha_i,\alpha,\alpha_r,\alpha_d$) and the spectral slope ($\beta$) all depend on $a,k,p$ (where the dependence on $a$ only arises during the ASDE phase). Thus, there are up to five (in case of a double peaked light-curve, three otherwise) observable slopes within a given synchrotron PLS. As multiple synchrotron PLSs could in principle be observable, the total number of observables could be much greater (up to five times the number of observed PLS segments). The number of potential observables far exceeds the number of free parameters, meaning that the underlying parameters are over-constrained. As a result, not only can $a,k,p$ be well constrained, the underlying model itself can be critically tested by such observations. In reality, in many cases it may be challenging to observationally probe all the phases of the lightcurve evolution, as they may extend over a wide range of timescales. Furthermore, as evident by GW170817, even if the lightcurve is observed for a long time, it is not trivial to observe multiple synchrotron PLSs. It is therefore of practical importance that even if an off-axis afterglow is observed only during the ASDE phase, and in a frequency range where the evolution of either $\nu_m$ or $\nu_c$ can be observed then this provides us with enough information to uniquely deduce $a,p,k$.

The other structure parameters: $b,\xi_{\rm c},q$ can be related to the ratios of characteristic timescales and fluxes. As shown by BGG20, for a double peaked lightcurve, all three parameters can be uniquely solved for from such observations. However, for a single peaked lightcurve only a single combination of these three parameters can be obtained and related to the start and end points of the ASDE phase: $t_{\rm pk}/t_{\rm dec}(\theta_{\rm F,0})$. 

Finally, we note that if $a<a_{\rm cr}$ (which may be appropriate for lGRB structures) the entire lightcurve evolution is changed, as the flux is dominated by material in the vicinity of the line of sight until late times when it becomes dominated by material further and further away from the jet core. As a result the flux and characteristic frequency temporal evolution completely changes. Furthermore, the flux centroid moves away from the core and has a more elongated shape compared to steep structures. Such a situation can be therefore clearly distinguished from a ``steep jet lightcurve" obtained for $a>a_{\rm cr}$. Therefore, the observation or lack of those features, provides an independent constraint on $a,p,k$ (see Table~\ref{tbl:acr}).

\subsection{Robust Features in Different Simulations}
When comparing 3D hydrodynamical simulations of jets propagating in expanding media from different works 
\citep[e.g.,][]{Gottlieb2021,Nativi2021,Lazzati+17a}, there is generally good agreement in the energy angular 
profile. Some differences in the outflow angular structure do emerge, most prominently in the angular profile 
of bulk $\Gamma$ as shown in the left panel of Fig.\,\ref{fig:model-profiles} and Table\,\ref{tab:model-fit-params}, 
with the conclusion that \textit{outflows that have a smaller $\Gamma_c$ have shallower power-law profiles}, 
i.e. smaller $b$. This is regardless of the initial setup that varies a lot between different simulations, 
and likely arises due to $\Gamma_0$ being only mildly relativistic at large angles because of baryon loading caused 
by mixing with the outer cocoon material. At the same time, $\Gamma_c$ approaches $\Gamma_\infty$ of the material 
that is injected at the base of the jet, and therefore the larger $\Gamma_c$ is the steeper (larger) is the resulting 
angular profile ($b$). 
When comparing the velocity angular structure of the lGRBs, the same conclusion could not be derived due to the 
small number of simulations presented here. When looking at Table 2 of \citet{Gottlieb2021}, there is no clear 
correlation between $\Gamma_c$ (or $u_\infty$, the maximum attainable proper velocity if all the jet pressure 
converted to kinetic energy, in their notation) of the lGRBs and the slope of the power-law profile.

\subsection{Magnetized Outflows}
In this work, we have only focused on the results of hydrodynamic (non-magnetized) jet simulations. 
When the magnetization of the flow
is raised significant differences in the properties of the two kinds of flows start to emerge. The 
principal difference is the suppression of mixing between the jet and cocoon material that leads to 
less energy being transferred to the interface between the two media \citep[e.g.,][]{Komissarov-99,Bromberg+14,Matsumoto-Masada-19,Gottlieb+20}. 
Even a modest magnetization, as low as $\sigma\sim10^{-2}$, is enough to significantly suppress this mixing. 
Consequently, the power-law angular profiles at $\theta>\theta_c$ are much steeper in comparison to the 
purely hydrodynamic case \citep[see, e.g.,][]{Nathanail+20,Nathanail+21}. It should be emphasized 
here that the angular structure is sensitive to the evolution of magnetization as the jet propagates and interacts 
with the confining medium, which depends on whether the magnetic energy dissipation, e.g. that occurs at collimation 
shocks, is well resolved in the simulation. This may be challenging to achieve in some simulations due to the 
large dynamical range, which may affect the final results.  
As demonstrated in Fig.\,\ref{fig:diffadiffk} 
and Eq.~(\ref{eq:Deltaalpha}), significant differences arise in the afterglow lightcurves when the energy 
angular profile becomes steeper. In particular, the lightcurve displays two peaks rather than one more readily, 
for a given viewing geometry. However, a competing effect is produced by steeper bulk $\Gamma$ angular profiles. 
From Fig.\,\ref{fig:1vs2}, it can be seen that for a given core compactness and viewing angle, steeper 
profiles more often lead to single peaks. Therefore, there is no obvious way to tell low and high $\sigma$ jets 
apart by simply looking at the afterglow lightcurve. Even in the case of GW\,170817, lightcurves derived from 
both hydrodynamic and MHD outflows were able to describe the afterglow equally well. Detailed modeling will 
be needed to ascertain the exact angular structure. Since magnetized jets have significantly steeper angular 
structures, it makes it difficult to observe the prompt emission from such misaligned outflows, i.e. 
$\theta_{\rm obs}/\theta_c$ must be close to or less than unity \citep{Beniamini-Nakar-19,Gill+20}. 

\section*{Acknowledgments}
PB's research was supported by a grant (no. 2020747) from the United States-Israel Binational Science Foundation (BSF), Jerusalem, Israel.
\section*{Data Availability}
The data produced in this study will be shared on reasonable request to the authors.

\bsp	
\label{lastpage}
\end{document}